\documentclass[prl,aps,amssymb,amsmath]{article}
\usepackage{amssymb,amsmath}

\begin{document}
\title{{\bf Index Theory and Adiabatic Limit in QFT}}
\author{Jaros{\l}aw Wawrzycki \footnote{Electronic address: jaroslaw.wawrzycki@wp.pl or jaroslaw.wawrzycki@ifj.edu.pl}
\\Institute of Nuclear Physics of PAS, ul. Radzikowskiego 152, 
\\31-342 Krak\'ow, Poland\\
\\
\ \ \ \ \ \ \ \ \ \ \ \ \ \ \ \ \ \ \ \ \ \ \ \ \ \ \ \ \ \ \ \ \ \ \ \ 
 \ \ \ \ \ \ \ \ \ \ \ \ \ \ \ \ \ \ \ \ \ \ \ \emph{To my wife}}
\maketitle
\newcommand{\ud}{\mathrm{d}}

\vspace{1cm}

\begin{abstract}
The paper has the form of a proposal concerned with the relationship between the three  mathematically rigorous approaches to quantum field theory: 1) local algebraic formulation of Haag, 2) Wightman formulation and 3) the  perturbative formulation based on the microlocal renormalization method. In this project we investigate the relationship between 1) and 3) and utilize the known relationships between 1) and 2). The main goal of the proposal lies in obtaining obstructions for the existence of the adiabatic limit (\emph{confinement} problem in the phenomenological standard model approach).  We extend the method of deformation of D\"utsch and Fredenhagen (in the Bordeman-Waldmann sense) and apply Fedosov construction of the formal index -- an analog of the index for deformed symplectic manifolds, generalizing the Atiyah-Singer index. We present some first steps in realization of the proposal.
\end{abstract}

\section{Introduction}

The paper has the form of a proposal concerned with the relationship between the three  mathematically rigorous approaches to quantum field theory. Namely they are: 1) local algebraic formulation of Haag, 2) Wightman formulation and 3) the  approach based on the microlocal causal renormalization method going back to Bogolubov and   St\"uckelberg, promoted mostly by Fredenhagen and his co-workers. In this project we investigate the relationship between 1) and 3) and utilize the known relationships between 1) and 2). The weakness of the $3)^{rd}$ approach lies in its dependence on the existence of the adiabatic limit, otherwise the formal power series are physically meaningless. With few exceptions only (e. g. QED) the existence problem for the adiabatic limit is open, even no  obstructions for its existence are known. The main goal of the proposal lies in obtaining obstructions for the existence of the adiabatic limit. The  problem corresponds to the \emph{confinement} problem in the phenomenological standard model approach. We extend the method of  deformation in the Bordeman-Waldmann sense as worked out by D\"utsch and Fredenhagen by noticing the parallelism between deformation applied by D\"utsch and Fredenhagen and the existence of adiabatic limit on the one side and the deformations of symplectic manifolds and the existence of the asymptotic representation of Fedosov on the other. It was suggested by Bordemann and Waldmann \cite{BorWal}. We extend their suggestion here. Fedosov constructed a formal analog of the index for deformed symplectic manifolds, generalizing the Atiyah-Singer index, and has shown that the existence of the asymptotic representation is equivalent to the integrality of the index. We notice further that the construction of his index may be applied to the D\"utsch and Fredenhagen deformations and  that his construction of necessity and sufficiency constraints may be carried to the  D\"utsch and Fredenhagen deformations provided we could utilize a Fredholm module over a fixed subalgebra of free fields, which is canonically connected to free fields. Quite independently we notice that  in the local algebraic theory the charges cannot superpose by principle, as they determine the selection sectors. Moreover in 1) there are two diverse kinds of non-superposing quantities, namely A) such as generalized charges and B) such as spacetime coordinates, i. e. classical  parameters with direct physical meaning, allowing the theory to have physical interpretation. We propose to treat them both more symmetrically in that the reason for the lack of coherent superpositions for B) should in principle be the same as for the lack of superpositions for A). Therefore B) should also be represented by the elements of the algebra of fields which do not mix the coherent selection sectors (of the Hilbert space acted on by all fields, also the charged fields)   and thus by elements which determine selection sectors. This leads us to  the concept of spacetime which is classical, i. e. whose points cannot superpose, but with noncommutative algebra of coordinates. In order to keep the geometric particle interpretation of Haag we identify the algebra with the Haag's algebra of detectors. Spacetime structure should determine its (pseudo-riemannian) spectral triple and (after “Wick rotation”) the corresponding Fredholm module. We identify the last module with the Fredholm module necessary for the construction (of the sufficiency condition) of the adiabatic limit. The first profit of this assumption is that it allows us to keep the particle interpretation even on curved spacetime without any time-like Killing vector field, a long standing problem in quantum field theory on curved spacetimes. Another profit:  we expect nontrivial limitations put on allowed values of coupling constants, which are deformation parameters in the D\"utsch-Fredenagen approach (integrality of Fedosov index assuring the existence of asymptotic representation  puts strong restrictions on possible values of deformation parameter). Last but not least we get the time's arrow for non-superposing quantities for free, as an immediate consequence of non-commutativity of multiplication in the algebra of spacetime coordinates. 

	The proposal is divided into five tasks: I. To provide details of the proof of the stability theorem under deformation of D\"utsch and Fredenhagen with the modification in definition of the algebra of observables meaning that we restrict ourselves to ghost-free fields in the construction of the algebra and explain the relationship between the two definitions (see section \ref{SQFT} for details). II. To reconstruct the asymptotic behavior of the analog of Fedosov asymptotic representation for QED utilizing the Blanchard-Seneor analysis and relationships between 1) and 2). III. To formulate necessary conditions for the existence of the asymptotic representation in QED in terms of the formal index. IV.  Having given a compact spectral triple to construct a formal deformation of the triple in the sense of Bordeman-Waldmann, and examine stability of the compact spectral triple structure under the deformation. V. Having given a completely integrable Faddeev model to investigate more deeply analytic properties of the linear representation of the quantum monodromy matrix on a dense subset of the Fock space, given in the Korepin, Bogoliubov and Izergin monograph (section \ref{time}). Then incorporating the relationship between point-like fields and local algebras try to carry the quantum group structure and their action on the corresponding spacetime algebra of bounded operators.

\section{A Tentative Hypothesis}\label{hyp}

In 1957 at the conference in Chapel Hill, Richard Feynman presented his famous \emph{Gedanken Experiment} supporting the claim, that the gravitational field has a quantum mechanical character in more or less the same sense as the electromagnetic field, and thus should be be quantized in more or less he same way as the electromagnetic field and other matter fields. The postulate that all physical processes (all the more quantum mechanical processes) should be described by amplitudes (and not probabilities themselves) was very natural at that time, i.e. only three decades after the discovery of matrix mechanics. Thus naively speaking: one can confine oneself to observables acting in a fixed Hilbert space and moreover there was no reason visible at that time for taking into consideration other collections (algebras) of observables than those which act irreducibly in the Hilbert space (Hermann Weyl in his famous book\footnote{See \cite{Weyl}, 238; moreover, he derives the Schr\"odinger equation   from the irreduciblility, compare 
\emph{Ibidem}, Chap. IV.D.} even referred to the Aristotelian \emph{nihil frustra} principle, in order to support the restriction to irreducible representations only). The mentioned postulate together with the "natural" assumption that, say, an electron is a spacetime object (in more or less the same sense as, say, a grain of sand) indeed gave a solid argument speaking for the quantum character of gravitational filed, \emph{i.e.} in the sense that it should undergo the superposition principle, and should be quantized similarly to matter fields.  The warning, which Feynman gave on that occasion, that quantum mechanics may not be correct for macroscopic objects, suggesting some possibilities for other alternatives has apparently at least been ignored, universally recognized as his scientific honesty at most. Nowadays, over half a century after the conference, the principal arguments of Feynman speaking for the quantum character of gravity, get lost much of their cogency. First of all the simplified scheme: observables + Hilbert space in which they act irreducibly, had to be substantially subtilized. According to the subsequent investigations in QFT and quantum statistical mechanics, we have all the grounds to expect, that the Hilbert space has to be divided into subspaces, called \emph{superselection sectors}, and the superposition of amplitudes cannot take place freely in the whole Hilbert space but only within one and the same sector, whenever the system in question is more complex.\footnote{Introduction of superselection sectors is not confined to high energy physics. For example it is an everyday practice of quantum chemists, e.g. there are two versions, namely left and right, of the thalidomide. As follows from the laboratory practice the two different versions do not superpose, but superpositions within one and the same version are possible. In fact the right version is effective against morning sickness in pregnant women. On the other hand the left specie produces foetal deformations, so that the possibility of superpositions between the two would have fatal biological consequences. Theory of quantum measurement is another example.} In particular it seems hardly possible that two states with different (generalized) charge numbers (e.g. different hadron numbers, or state with even half spin with a state with an odd half spin) can superpose. Beyond doubt the assumption that such states cannot superpose is a "good approximation" in the light of nowadays knowledge, independently of the possible disputes on how is in "reality". Similarly we have all the grounds to suppose, that states with different electric charge numbers do not superpose (although there exists in this case an alternative theory of  A. Staruszkiewicz, compare discussion below).\footnote{It should be emphasized here that this is an assumption (or hypothesis) of experimental tentative character and it cannot be mathematically inferred from the ordinary quantum mechanics contrary to what is sometimes misstated. All "proofs" of the theorems "that such and such quantity is classical in the sense that it does not undergo the superposition principle" turned up ultimately to be ineffective and contained serious gaps, e. g. that the argument of \cite{WicWigWig} falls short of the claim was subsequently shown in \cite{AhaSus}. Essentially on the same footing the "proof" of Lifshitz and Pitaevski, that the Coulomb field is classical as well as other similar "proofs" fall short of their goals because the superselection structure goes beyond the competencies of the ordinary quantum mechanics.}         From the investigations of Haag and his school \cite{Haag} (algebraic quantum field theory) it follows that the charge structure (\emph{global} gauge symmetry groups) can be obtained from the structure of the equivalence classes of representations of the algebra of observables, where all the representations in question come out of a special natural class fulfilling the so called \emph{superselection condition}.\footnote{One can reconstruct in this way e. g. the isospin group. However some subtle difficulties arise in case of the electric charge in choosing the suitable selection rule and the suitable class of representations in this case. They are caused by the unlimited range of the electromagnetic interactions (zero rest mass of the photon) and with the construction of the "Hilbert space" with the indefinite product – within the algebraic formalism it is difficult to construct such space and to distinguish the Hilbert space of "\emph{physical states}" in it. Below we return to this problem.}    The algebra of (quasi-local) observables has the property (among other properties) that transforms sector into the same sector (of the Hilbert space of point-like fields in the sense of Wightman, corresponding to the algebra of quasi-local observables, in which all the point-like fields -- also the charge carrying fields with nontrivial gauge -- act, whenever such corresponding fields do exist\footnote{Compare discussion below; the relation between quasi-local algebra in the sense of Haag and pointlike fields in the sense of Wightman is essential for the whole proposal.}\label{correspondence}). Thus no element of the observable algebra\footnote{More precisely: no Wigtman point-like filed smeared out over a compact domain, corresponding to an element of the local algebra of observables leads us out of the superselection sector.} leads us out of a coherent subspace (selection sector). Roughly speaking the superselection condition allowing us to select the natural class of representations tells that in space-like infinity each representation of the class behaves like the vacuum representation (there are some important troubles\footnote{From time to time opinions arise claiming that the troubles have only technical character and are not fundamental, compare e.g. \cite{Rob}. Some knew perspectives was presented in: \cite{BuchDop}. But this \emph{status quo} lasts since the early seventies of the previous century without any breakthrough visible in the solution. It seems that it will be difficult to avoid the analysis of the relation of the algebraic (Haag's) and Wightman formulation with the perturbative formulation of QFT, compare \cite{BuchDop, Buch, BruFred}.} just with this condition for the electric charge). This suggests that the quantities which do not undergo the superposition principle, such as charges, are characterized by a decomposition parameters of representations of the observable algebra (or some subalgebra of the corresponding algebra of smeared out fields, whenever they exist) into irreducible representations. Roughly but suggestively speaking: non-superposing quantities are decomposition parameters of the representation of the smeared out fields corresponding to observables or some other distinguished subalgebra of fields into irreducible representations. Similarly in the quantum statistical mechanics the quantities which do not superpose shows up as decomposition parameters of representations of the same algebra into irreducible representations, but this time for representations of the statistical and not the vacuum sector. Haag's approach and his school partially based on observable algebra understood in the classical sense (introduced by Dirac in his famous handbook on quantum mechanics) but taking into account division into selection sectors, thus based on representation theory of one and the same algebraic structure, was not able as yet to explain in the same manner the structure of \emph{local} gauge group symmetries; not to mention the difficulties with electric charge and indefinite product. In my opinion the fundamental reason for the lack of success here lies in this: The algebraic theory introduces two kinds of non-superposing quantities with no deeper interrelation between them: 1) such as charges and 2) such as spacetime coordinates, which are classical quantities with direct physical interpretation, enabling  the algebraic theory to have a physical interpretation, yet the the local gauge symmetries connect the two kinds of quantities. Here I propose the

\vspace{0.5cm}

POSTULATE. \emph{Not only generalized charges, but all non-superposing quantities, including classical directly observable parameters, should be decomposition parameters of representations of some fixed subalgebra of the algebra of smeared out point-like fields into irreducible representations}.

\vspace{0.5cm}

This is of course a hypothesis of tentative character. In order to keep the physical interpretation and in order to enable concrete computations, we have to supply the postulate and have to point out the subalgebra which corresponds to the algebra of spacetime coordinates. Namely we supply the postulate with the hypothesis that the subalgebra is given by the so called *-algebra of detectors  $\mathcal{A}$ (not unital; roughly speaking it is generated by the elements of the observable algebra of the form $L^* L$, where $L$ are quasi local annihilators, which differ from the Doplicher annihilators only by the property that the ideal which they form is not norm closed)\footnote{Compare \cite{Haag}, p. 283, algebra of detectors is denoted there by $\mathcal{C}$.}. Therefore we admit the classical quantities to possess their own subalgebra, which determines their own superselection sectors. At the preliminary stage at least it seems reasonable to assume that the algebra of spacetime coordinates determines the universal structure of superselection sectors for all macroscopic quantities (compare the geometric physical interpretation proposed by Haag \cite{Haag}). The physical motivation for this definition follows from the geometric particle interpretation of the algebraic theory proposed by Haag as well as the role of classical  spacetime coordinates and the algebra  of detectors (Doplicher annihilators) in this interpretation \cite{Haag}.  Actually a similar postulate was put foreword by Haag himself, when expressing the conjecture that \emph{local} gauge groups can be explained within the algebraic formulation similarly as the \emph{global} groups, allowing a wider class of representations of observable algebra, in particular going out of all the sectors of the Hilbert space in which the algebra of corresponding (smeared out) point-like fields (including charge carrying fields) acts. Haag's postulate, however, does not give any explicit computational hints (which, among other things, is confirmed by the lack of its realization); in particular it is not clear how to look for the additional representations. Such additional representations would be necessary if no other subalgebra besides the observable algebra would be allowed to determine the superselection sectors. Although some restriction of this kind has to be put on the allowed subalgebra in order to give an objective sense to a non-superposing quantity, the algebra of observables is too big and we have to seek a smaller one. If no other algebra fixing the superselection sectors than the algebra of observables were allowed, no \emph{local} superselction  sectors would be left, by the \emph{local normality principle}\footnote{\cite{Haag}, p. 131.}. However there is no indication (within the algebraic theory) that the algebra of observables determines all relevant selection sectors, e. g. all sectors sufficient to define all relevant non-superposing macroscopic quantities, sufficient for the physical interpretation of the theory. Even contrary: for the geometric particle interpretation at least all information comes from the use of the subalgebra of detectors and coincidence arrangements of detectors\footnote{\cite{Haag}, p. 272.}, indicating that the subalgebra of detectors is sufficient to pick up all relevant sectors, thus suggesting that the whole algebra of observables mix too many sectors of objective physical meaning. Moreover the assumption that non-superposing quantities (including macroscopic quantities) should be construable via the selection sectors \emph{inside} the Hilbert space acted on by the corresponding\footnote{Whenever such fields do exists and the correspondence mentioned in the footnote 5 is meaningful, compare discussion below.} (smeared out) point-like fields (exactly as for the the charges and the algebra of observables) finds a justification in the fact that the quantum theory of fields is in agreement at least with the phenomenological theory\footnote{ I mean the well known FAPP-type methods of H. \.Zurek and his school.} of quantum measurement, assuming that the detectors determine their own selection sectors (assumption which cannot be derived from the ordinary quantum mechanics, as was emphasized e. g. by Penrose).  
  
	Perhaps we should explain that the \emph{classical} character of spacetime  (in the physical sense used here) and the \emph{non-commutative} character of the algebra of spacetime coordinates ($\mathcal{A}$)  are not \emph{a priori} inconsistent. The term \emph{classical} used here means that the superposition takes place only within one and the same sector (of the sector structure in the Hilbert space acted on by the fields, determined by the smeared out fields corresponding to $\mathcal{A}$). Therefore no superposition exist between (elements of spaces of) in-equivalent irreducible representations of $\mathcal{A}$; thus no "superposition" of two different spacetime "points" can exist; as the "points" of $\mathcal{A}$ correspond to equivalence classes of irreducible representations of $\mathcal{A}$.\footnote{Therefore the "parameters" numbering irreducible representations of the spacetime algebra cannot superpose. In passing: also the classical manifold (in the sense: commutative) can be described by a non-commutative algebra Morita equivalent to the commutative algebra of smooth functions on the manifold, compare e. g. \cite{Connes}. Of course this case is trivial from the physical point of view, and by this, it is not very interesting for physicists. }  Parameters numbering the irreducible representations are  in a one-to-one correspondence with the spectrum of a commutative subalgebra $\mathcal{A}_{cl}$. Assume for a moment (only for heuristic aims) that  $\mathcal{A}_{cl}$ is a subalgebra of  
$\mathcal{A}$ and therefore it is equal to the center of $\mathcal{A}$.\footnote{In general for decompositions into irreducible representations  $\mathcal{A}_{cl}$ is a maximal commutative subalgebra in the commutant of $\mathcal{A}$. Here we assume that the algebra $\mathcal{A}$ acts in a fixed Hilbert space, the same in which the corresponding Wightman fields act, and assume that the action defines a faithful representation of $\mathcal{A}$ which is to be decomposed.} We can therefore localize $\mathcal{A}$ with respect to $\mathcal{A}_{cl}$. Heuristically the elements $A(x)$ of the localization with $x$ ranging over open subsets $\mathcal{U}$ of the spectrum of $ \mathcal{A}_{cl}$ are the elements of the algebra of detectors $\mathcal{A}(\mathcal{U})$ on\footnote{This is only heuristic, as detectors are localizable only asymptotically.} $\mathcal{U}$. This "approximation" is, however, too coarse and unrealistic. In the geometric particle interpretation at least, we consider detectors (asymptotically) localized within compact subsets. Although the subsets are small in comparison to distances between localization centers and two detectors with different localization centers (asymptotically) commute, in all relevant coincidence arrangements of detectors, see \cite{Haag} p. 272, they cannot be shrunk to \emph{points}. Here \emph{points} are used in ordinary commutative sense, and have immediate physical meaning of spacetime points used in algebraic  quantum field theory (which we intend to identify with elements of the spectrum of $\mathcal{A}_{cl}$). Therefore we are forced to use a non-commutative localization, say of Ore type, with respect to a commutative subalgebra $\mathcal{A}_{cl}$ of $\mathcal{A}$ not contained in the center of $\mathcal{A}$. Of course these are heuristic remarks only, motivated on the geometric interpretation of Haag, suggesting that in general realistic situation there should exist a commutative subalgebra $\mathcal{A}_{cl}$ in the algebra of detectors $\mathcal{A}$ whose spectrum elements are parameters with immediate physical interpretation\footnote{In general we cannot, however, expect that the spectrum of $\mathcal{A}_{cl}$ will be sufficient to designate all points of $\mathcal{A}$, for example the representation of $\mathcal{A}_{cl}$  induced by an irreducible representation of $\mathcal{A}$ is not irreducible in general if $\mathcal{A}_{cl}$ does not lie in the center of $\mathcal{A}$. Algebraically speaking:  possibly many different localizations are needed to reconstruct the algebra $\mathcal{A}$ and its relevant spectrum,  giving different types of coincidence arrangements of detectors. Most of all we should be interested in the coincidence arrangements of detectors encountered in particle physics, of course.}. In order to check the consequences of the above postulate (suitably supplemented) one have to introduce (natural) analytic structures allowing concrete computations. We shell describe only some first steps towards this goal, based on the  (rigorous) micro-local perturbative approach of Brunetti and Fredehagen and formulate its connection to local algebraic approach of Haag in terms of formal index theorem of Fedosov and asymptotic representations (generalizing the asymptotic representations of Fedosov). This allows us to introduce spectral triple formalism of Connes, via its construction for free fields.

\section{Spacetime and QFT}\label{SQFT}

	Here we formulate the hypothesis of the previous section in more concrete mathematical terms. We use the tools of non-commutative geometry, and introduce a natural structure of spacetime in terms of this geometry, which may be adopted to this operator-algebraic situation and explain its natural connection to structures which one finds in quantum field theory.  We will use the local perturbative construction of the algebra of observables in gauge theories as proposed by D\"utsch, Fredenhagen and Brunetti \cite{DutFred, BruFred}. But first we remind that the analog\footnote{ There are several competitive proposals for this analogue, some of them propose to include (the smooth) fundamental symmetries into the construction of the operator $D$ acting in ordinary Hilbert space (for example Connes and Marcolli \cite{ConMar} propose to construct a spectral triple in ordinary Hilbert space abandoning, however, (ordinary) self-adjointness of $D$, but keeping the self-adjointness of $D^2$). We rejected those propositions whose construction is based on foliations into Cauchy hyper-surfaces, which seem to be less general. The non-compact riemannian case (non-unital $\mathcal{A}$) is worked out in: \cite{Gay}. Actually first steps has been prepared only in this non-compact direction, but no fundamental difficulties are expected here. An extension of spectral triple formalism to type III algebras has been proposed in: \cite{ConMosIII}. }      $(\mathcal{A}, D, \mathcal{H})$ of the Connes' spectral triple for pseudo-riemannian manifold, as proposed by Strohmaier \cite{Stro}, is given by a pre-C*-algebra $\mathcal{A}$ with involution * acting as an algebra of bounded operators not in the ordinary Hilbert space but in a Krein space\footnote{Let us remind  briefly that the Krein space \cite{Bog} 
$\mathcal{H}$ is a linear space with indefinite non-degenerate inner product $(\cdot , \cdot)$ which admits a direct sum decomposition $\mathcal{H}_+ \oplus \mathcal{H}_-$  into subspaces $\mathcal{H}_+$ and $\mathcal{H}_-$  on which $(\cdot , \cdot)$ is positive definite and respectively negative definite and such that $\mathcal{H}_+$ and $\mathcal{H}_-$  are closed in norm topology induced on them by the inner product $(\cdot , \cdot)$. Thus $(\cdot , \cdot)$ induces on $\mathcal{H}_+$ and $\mathcal{H}_-$  the structure of ordinary Hilbert spaces. For any such decomposition $\mathcal{H}  =  \mathcal{H}_+ \oplus \mathcal{H}_-$ , one defines  the operator of fundamental symmetry $\mathfrak{J}$ putting it equal to  +1 on $\mathcal{H}_+$ , and   $-1$ on $\mathcal{H}_-$. Moreover $\langle \cdot , \cdot \rangle_\mathfrak{J}$  =  $( \cdot , \mathfrak{J} \cdot)$ is an ordinary positive definite inner product inducing on $\mathcal{H}$  an ordinarily Hilbert space structure. Norms induced by these inner products  defined by any two symmetries $\mathfrak{J}$ are equivalent.}\label{Krein}        \cite{Bog} $\mathcal{H}$. The involution is represented by taking the Krein adjoint, the Dirac operator $D$ is self-adjoint in the Krein sense. Important role is played by the so called fundamental symmetries of the Krein space $\mathcal{H}$. These are operators: $\mathfrak{J} : \mathcal{H} \to \mathcal{H}$ , such that:  $\mathfrak{J}^2  =  1$  and   $(\cdot, \mathfrak{J} \cdot)$ = $(\mathfrak{J} \cdot, \cdot)$, where $(\cdot, \cdot)$ is the indefinite inner product in the Krein space $\mathcal{H}$. With the help of $\mathfrak{J}$, one can obtain ordinary (riemannian) spectral triples from pseudo-riemannian spectral triples in a similar way as this is done in quantum field theory by "Wick rotation", when passing to riemannian signature. After this digression we go back to the perturbative construction of the algebra of observables as proposed in \cite{DutFred, BruFred}. We start from free fields in a theory with gauge symmetry. Afterwards we construct the algebra of fields (and algebras of observables and detectors) without performing the adiabatic limit, noticing that their construction depends locally on interaction. It is based on the old ideas of Bogoliubov and St\"uckelberg, developed by Epstein and Glaser, and then by 
D\"utsch, Brunetti and Fredenhagen, who applied to it the H\"ormander's microlocal analysis of wave fronts for hyperbolic operators. The prise we pay for the clear separation of \emph{local} aspects (renormalization) from the \emph{global} (adiabatic limit) lies in this: algebras thus constructed are formal power series algebras only, with mathematically well defined coefficients built of free fields, and are deformations of the free filed algebras in the sense of Bordemann-Waldmann \cite{BorWal}. Therefore only a halfway is thus reached: the existence of adiabatic limit remains to be examined. We return to the existence problem below, but first we give some details of the construction of D\"utsch and Fredenhagen. The local algebra $\mathcal{F}(\mathcal{U})$ of free fields with gauge symmetry (as well as interacting fields, if one assumes the existence of adiabatic limit) does not act in ordinary Hilbert space, but in a space with indefinite inner product, compare the Gupta-Bleuler formalism. In order to give a mathematical sense to some operator manipulations performed by physicists some assumptions of topology-analytic character are necessary (to make the various kinds of taking adjoint of an operator more precise, \emph{etc.}). We assume in particular that $\mathcal{H}$ is a Krein space (indefinite inner product is non-degenerate and the subspaces $\mathcal{H}_+$ and $\mathcal{H}_-$  of the footnote 18 are closed in norms induced by the indefinite inner product).  Thus the Gupta-Bleuler operator\footnote{This operator was denoted by $\eta$ in the Polish translation of the book: W. Heitler, {\sl "The Quantum Theory of Radiation"}, Clarendon Press, Oxford, 1954, §II.10.}           $\eta$ is a fundamental symmetry of the Krein space $\mathcal{H}$ (one of many such, and which was denoted above by $\mathfrak{J}$). It is clear that also in this situation we can repeat the general argument of Haag, that the elements of the algebra of fields which represent observables cannot lead us out of coherent subspaces of $\mathcal{H}$. This time, however, situation is more complicated, as we identify two vectors of $\mathcal{H}$ which differ by the so called "admixture", a vector on which the indefinite inner products is zero and, moreover, not all vectors  of $\mathcal{H}$  are regarded as physical (in particular the indefinite inner product must be positive on them). In order to reconstruct the so called physical Hilbert space $\mathfrak{H}$ we have to use the full BRST formalism (or its equivalent, D\"utsch and Fredenhagen use the Kugo-Ojima operator $Q$). In particular the net $\mathcal{U} \to \mathcal{F}(\mathcal{U})$  of local fields is such that every local algebra $\mathcal{F}(\mathcal{U})$  is a  *-algebra with a $\mathbb{Z}_2$-gradation. A graded derivation $s$ acts on the algebra $\mathcal{F}$ = $\bigcup_{\mathcal{U}} \mathcal{F}(\mathcal{U})$  of quasi-local fields, such that $s^2 = 0$, $s(\mathcal{F}(\mathcal{U})) \subseteq \mathcal{F}(\mathcal{U})$, $s(F^{*}) =  -(-1)^{\delta(F)} s(F)^*$,  $s(AB)= s(A)B + (-1)^{\delta(A)}As(B)$,  where the  $\mathbb{Z}_2$-gradation is defined by $F \to (-1)^{\delta(F)} F$   ($\delta(F)$ is the ghost number of the field $F$, and $s$ is the  BRST transformation). From the properties of $s$ it follows that the kernel ker$s$ = $\mathfrak{A}_0$ as well as the image  $s(\mathcal{F}) = \mathfrak{A}_{00}$ of derivation $s$ are *-sub-algebras of 
$\mathcal{F}$. D\"utsch and Fredenhagen define then the algebra of quasi-local observables and the net of local observables as follows:
\begin{displaymath}
    \mathfrak{A}_0 \, \textrm{mod} \, \mathfrak{A}_{00} \,\,\,\,\,\,\,\,\,\,\,\,
 \textrm{and} \,\,\,\,\,\,\,\,\,\,\,\, \mathcal{U} \, \longrightarrow \, \mathfrak{A}_0 \cap \mathcal{F}(\mathcal{U}) \,\,\, \textrm{mod} \,\,\, \mathfrak{A}_{00} \cap \mathcal{F}(\mathcal{U}),  
\end{displaymath}
which makes sense because $s^2 = 0$ and $\mathfrak{A}_0  \supseteq \mathfrak{A}_{00}$. The action of field operators on $\mathcal{H}$ is such that the involution is represented by the Krein adjoint. We assume additionally that the gradation on $\mathcal{F}$ can be represented by a $\mathbb{Z}_2$-gradation on $\mathcal{H}$, such that $\mathfrak{A}_{0}^{(0)}$, $\mathfrak{A}_{00}^{(0)}$ and  $\mathfrak{A}_{0}^{(1)}$ , $\mathfrak{A}_{00}^{(1)}$  are  subspaces in   $\mathfrak{A}_0$    and $\mathfrak{A}_{00}$  of grade 0 and 1 respectively. We adopt this gradation as the gradation of the (even) pseudo-riemannian spectral tiple $(\mathcal{A}, D, \mathcal{H})$ mentioned above. We propose also a slight modification in the above definition of the algebra of observables (quasi-local and local) and we put  instead:
\begin{displaymath}
\mathfrak{A} = \mathfrak{A}_{0}^{(0)} \, \textrm{mod} \, \mathfrak{A}_{00}^{(0)} \,\,\,\,\,\,\,\,\,\,\,\,
 \textrm{and} \,\,\,\,\,\,\,\,\,\,\,\, \mathcal{U} \, \longrightarrow \, \mathfrak{A}(\mathcal{U}) = \mathfrak{A}_{0}^{(0)} \cap \mathcal{F}(\mathcal{U}) \,\,\, \textrm{mod} \,\,\, \mathfrak{A}_{00}^{(0)} \cap \mathcal{F}(\mathcal{U}),  
\end{displaymath}
thus confining ourselves in their definition to fields with even ghost number. The algebra $\mathcal{A}$ of spacetime coordinates is not directly identified with the algebra of detectors, but with the sub-algebra $\mathcal{A}$ of $\mathfrak{A}_{0}^{(0)}$ for which $\mathcal{A}$ mod $\mathfrak{A}_{00}^{(0)}$ is the algebra of detectors (this is the identification proposed above with the necessary modification caused by the fact that not all vectors of the Krein space $\mathcal{H}$ are physical and by the identification of vectors differing by an "admixture"). We construct the representation of the algebra of observables in the ordinary (physical) Hilbert space
$\mathfrak{H}$  exactly as D\"utsch and Fredenhagen. If the graded commutator with an operator $Q$ represents $s$ (in short: if $Q$ represents $s$), \emph{i. e.}
\begin{displaymath}
s(F) =  QF  - (-1)^{\delta(A)} FQ, 
\end{displaymath}        
then $Q$ has to be self-adjoint in the sense of Krein and $Q^2 = 0$ (in order to ensure fulfillment of the following conditions $s(F^{*}) =  -(-1)^{\delta(F)} s(F)^{*}$ and $s^{2} = 0$).  Because the physical vectors should be  $s$-invariant D\"utsch and Fredenhagen introduce the following definitions: $\mathcal{H}_{0}$ = ker $Q$ and $\mathcal{H}_{00}$ = Im $Q$.  Then they assume:

\begin{enumerate}

\item[(i)]
            $(\varphi, \varphi) \geq 0$ for every $\varphi \in \mathcal{H}_{0}$    
           ({\bf Positivity}),     
\item[(ii)]
            [$\varphi \in \mathcal{H}_{0} \land  (\varphi, \varphi) = 0$] 
             $\Longrightarrow$ $\varphi \in \mathcal{H}_{00}$;

\end{enumerate}

 and put

\begin{displaymath}
\mathfrak{H} = \mathcal{H}_0 \, \textrm{mod} \, \mathcal{H}_{00},   
\end{displaymath} 

with the following inner product on $\mathfrak{H}$ : 
  
\begin{displaymath}
\langle [\varphi_{1}], [\varphi_{2}] \rangle❭_{\mathfrak{H}}  =  (\psi_{1} , \psi_{2}), \,\,\,\,\,\,\, \psi_{j} \in [\varphi_{j}]= \varphi_{j} + \mathcal{H}_{00}.
\end{displaymath}

$\mathfrak{H}$ with so defined inner product is a pre-Hilbert space (the closure turns $\mathfrak{H}$   into a Hilbert space). Next, the formula
\begin{displaymath}
\pi([A])[\varphi] = [A\varphi], \,\,\,\,\,\,\,\textrm{where} \,\,\,\, 
[A] = A + \mathfrak{A}_{00}^{(0)} \,\, \textrm{and} \,\,    A \in \mathfrak{A}_{0}^{(0)}, 
\end{displaymath}
define a *-representation of the algebra of observables with involution * represented by the ordinarily adjoint. D\"utsch and Fredenhagen consider only the case $[A] = A + \mathfrak{A}_{00}$,    $A \in \mathfrak{A}_0$. Next we prove after D\"utsch and Fredenhagen that the construction of the algebra of observables and its representation is stable under Bordemann-Waldmann deformatons, \emph{i. e.} after the interaction is "switched on". Thus if one starts from free fields acting on the Krein space 
$\mathcal{H}$ , and then construct the deformation of the algebra of fields, 
\emph{i. e.} build the formal power series of free fields  via the microlocal method of Brunetti-Fredenhagen \cite{BruFred}, then one can extend naturally the above construction of representation of observables (and detectors) for free fields to a formal Bordeman-Waldmann-type representation of deformed algebras of observables and detectors. To formulate strictly the "stability" theorem we need to introduce some further definitions. Namely in order to construct the deformation we replace every element $F \in \mathcal{F}$ with a formal power series\footnote{It is not important here but in computational practice, i. e. in the formal power series of the microlocal renormalization method of Brunetti-Fredenhagen, $g$ is a smooth function on the spacetime manifold (understood in the ordinarily sense) with compact support -- local coupling "constant".}  $\tilde{F}  = \sum_{n} g^{n} F_n$,  in which $F_0 = F$, $F_{n} \in \mathcal{F}$, $\delta(F_{n}) = const.$ We replace $s$ and $Q$ with a similar power series  $\tilde{s}= \sum_{n} g^{n} s_n$ (every $s_n$ is a graded derivation), $\tilde{Q} = \sum_{n} g^{n} Q_n$,  $Q_{n} \in \mathcal{L}(\mathcal{H})$, $s_{0} = s$, $Q_{0} = Q$, thus
\begin{displaymath}
\tilde{s}^2 = 0, \,\, \tilde{Q}^2 = 0, \,\,  (\tilde{Q}\phi, \psi) 
= (\phi, \tilde{Q}\psi),
\,\, \tilde{s}(\tilde{F}) 
= \tilde{Q}\tilde{F} - (-1)^{\delta(\tilde{F})} \tilde{F}\tilde{Q}.
\end{displaymath}

Next we define the formal algebra of observables\footnote{M. D\"utsch and K. Fredenhagen define here: (ker$\tilde{s}$)   mod  (Im$\tilde{s}$). }      
by (ker $\tilde{s})^{(0)}$  mod    (Im$\tilde{s})^{(0)}$, and then replace  $\mathcal{H}_0$ and $\mathcal{H}_{00}$   with   $\tilde{\mathcal{H}}_0$ = ker$\tilde{Q}$  and    $\tilde{\mathcal{H}}_{00}$ = Im$\tilde{Q}$ and define  $\tilde{\mathfrak{H}}$   = ker$\tilde{Q}$ mod Im$\tilde{Q}$. The inner product in $\mathfrak{H}$  induces an inner product in $\tilde{\mathfrak{H}}$  which assumes values in a formal power series field over $\mathbb{C}$. It follows from the above construction that the formal algebra of observables has a natural formal representation $\tilde{\pi}$  on $\tilde{\mathfrak{H}}$.  D\"utsch and Fredenhagen adopt the definition that a formal power series $\tilde{b} = \sum_n g^n b_n$, $b_n \in \mathbb{C}$, is \emph{positive} iff there exists another power series $\tilde{c} = \sum_n g^n c_n$, $c_n \in \mathbb{C}$, such that $\tilde{c}^{*}\tilde{c} = \tilde{b}$, \emph{i.e.} such that $b_n = \sum_{k=0}^{n} \bar{c_n}c_{n-k}$. 

	In this situation D\"utsch and Fredenhagen proved the following stability theorem under deformation\footnote{With a slightly different definition of 
the algebra of observables, as has already been indicated above.}: If the positivity assumption is fulfilled, then 

\begin{enumerate}

\item[(i)]
            $(\tilde{\varphi}, \tilde{\varphi)} \geq 0$ for every $\tilde{\varphi} 
           \in   \tilde{\mathcal{H}}_{0}$    
               
\item[(ii)]
            [$\tilde{\varphi} \in \tilde{\mathcal{H}}_{0} \land  
             (\tilde{\varphi}, \tilde{\varphi}) = 0$] 
             $\Longrightarrow$ $\tilde{\varphi} \in \tilde{\mathcal{H}}_{00}$;
\item[(iii)]
            For every $\varphi \in \mathcal{H}_0$  there exists $\tilde{\varphi} \in
           \tilde{\mathcal{H}}_{0}$, such that $(\tilde{\varphi})_0 = \varphi$.
\item[(iv)]
              Let $\pi$ and $\tilde{\pi}$ be the representation of free field algebra constructed above and the formal representation of its deformation in $\mathfrak{H}$  or $\tilde{\mathfrak{H}}$ respectively. Then 

\end{enumerate}
\begin{displaymath}
\tilde{\pi}(\tilde{A}) \neq 0 \,\,\, \textrm{if} \,\,\, \pi((\tilde{A})_0) \neq 0.  
\end{displaymath}

\vspace*{1cm}

A state $\omega$ on the algebra of observables $\tilde{\mathfrak{A}}(\mathcal{U})$ is defined by the following conditions (compare \cite{DutFred, BorWal})

\begin{enumerate}

\item[(i)]
             $\omega$: $\tilde{\mathfrak{A}}(\mathcal{U}) \to$ $\tilde{\mathbb{C}}$ is
             linear, \emph{i. e.} $\omega(\tilde{a}[\tilde{A}] + [\tilde{B}])$ 
             = $\tilde{a}\omega([\tilde{A}])$ + $\omega([\tilde{B}])$,

\item[(ii)]
             $\omega([\tilde{A}]^{*}) = \omega([\tilde{A}])^{*}$ for all 
             $[\tilde{A}] \in \tilde{\mathfrak{A}}(\mathcal{U})$, 

\item[(iii)]
             $\omega([\tilde{A}]^{*}[\tilde{A}]) \geq 0$ for all
             $[\tilde{A}] \in \tilde{\mathfrak{A}}(\mathcal{U})$ and

\item[(iv)]
             $\omega(\tilde{{\bf 1}}) = \tilde{1}$.

\end{enumerate}

The physical vector-states constructed in \cite{DutFred} define naturally states in the Bordemann-Waldmann sense \cite{BorWal}:
\begin{displaymath}
\omega_{\tilde{\phi}}([\tilde{A}]) 
= \langle [\tilde{\phi}], [\tilde{A}][\tilde{\phi}] \rangle_{\tilde{\mathfrak{H}}},
\,\,\,\,\, [\tilde{\phi}] \in \tilde{\mathfrak{H}},
\end{displaymath}
where positivity follows from the positivity of Wightman distributions of gauge invariant fields, see \cite{DutFred}.

\vspace*{0.5cm}

\begin{tabular}{|p{11.2cm}|}\hline
We have thus arrived at the first preliminary task of our proposal: to provide 
details of the proof of the \emph{stability theorem under deformation} formulated above with the modifications indicated (\emph{i. e.} with the modification in definition of observable algebra and explain relationship between the two definitions). \\ \hline
\end{tabular}

\vspace*{0.5cm}

Actually the first part of this task follows from the proof of the stability theorem as presented in \cite{DutFred}, because $s$ preserves gradation. Only the comparision of the two definitions of the algebra of observables needs a closer inspection, but again, the relation between the two definitions for the free theory undrlying QED may essentially be read of from \cite{DutFred}. In this case the representation $\pi$ of our algebra of observables constructed above, in contrary to the algebra of observables of D\"utsch and Fredengagen, \emph{is faithful}, and it is generated by $[F^{\mu \nu}], [\psi], [\bar{\psi}]$ and Wick monomials thereof (of course here $[\cdot]$ are understood as classes modulo elements of the ideal $\mathfrak{A}_{00}^{(0)}$), whereas the "canonical" representatives of $\mathfrak{H}$ are vectors (of $\mathcal{H}$) containing transversal photons, electrons and positrons only, as follows from \cite{DutFred}. Our definition of the algebra of observables is therefore justified in
the free theory underlying QED at least and we can in this case confine ourselves to fields with even ghost number when constructing observables. What remains to be investigated in the first task is the relation between the two definitions of the algebra of observables for theories with more involved gauge freedom.

	Now we pass to the existence problem for the adiabatic limit, which in the formulation of  D\"utsch and Fredenhagen is equivalent to the following question: under what (accessible) conditions the formal seres are convergent, and thus when the formal representation of the deformed algebra turns into an actual representation of an actual (C*-)algebra in an ordinarily Hilbert space? But on the other hand Fedosov \cite{Fed} proved an interesting theorem in the theory of deformations of symplectic (or even Poisson) manifolds. Namely he showed that the deformed algebra admits a so called asymptotic operator representation in ordinary Hilbert space iff his (Fedosov') formal index fulfills some integrality conditions. His formal index is a formal analog of the Atiyah-Singer index (better: it is a generalization of the Atiyah-Singer index adopted to deformed algebras and their formal representations), in particular it is a topological invariant of the symplectic manifold, so it is an invariant of the algebra (of smooth functions on the manifold), which is subject to deformation as well as of the deformed algebra  (the latter in the general non-commutative sense: it is a formal K-theory invariant). The formal (Fedosov') index can be carried on deformations considered here. The  algebra of free fields (or rather the non-commutative algebra $\mathcal{A}$ of spacetime defined above, corresponding to free fields) plays the role of the algebra of smooth functions on the symplectic manifold subject to deformation. Next we confine ourselves to the QED case (in the above deformation formulation, compare e. g. \cite{DutFred}). We know that in this case the adiabatic limit does exist, i.e. Wightman distributions do exist (or Green functions), according to the Blachard-Seneor \cite{BlaSen} theorem. We may therefore reconstruct the action of smeared out fields (a construction by now rather well known formally analogous to the Gelfand-Naimark-Segal construction of representation from a state, firstly applied by Wightman). Having given this and the machinery of constructing local algebras (of bounded operators) from fields \cite{DriFro} we intend to read of the asymptotic conditions fulfilled by the representation so constructed which are induced by the asymptotic conditions of Blanchard and Seneor's paper, fulfilled by Green functions. We may thus construct the analog of Fedosov' asymptotic representation with the explicit asymptotic conditions fulfilled by power series of which we \emph{a priori} know that they admit an actual representation.  

\vspace*{0.5cm} 

\begin{tabular}{|p{11.2cm}|}\hline
Thus we arrive at the second task of our proposal: to formulate necessary conditions for the existence of the asymptotic representation of deformation in QED in terms of the formal index.\\ \hline
\end{tabular}

\vspace*{0.5cm}

To this end we intend to mimic the argument of Fedosov which he applies in the construction of the analogous necessary condition\footnote{Compare the theorem 7. 1. 2 and its proof in ref. \cite{Fed}.}: just as in the case of Fedosov's necessary conditions we expect that they will ultimately depend on (ordinary) K-theory invariant of the algebra subject to deformation (in our case this is the algebra $\mathcal{A}$ for free fields and its representation constructed as above). We expect to obtain in this way integrality-type conditions for the index on $\mathcal{A}$ (for free fields) which we propose to compare with the index map induced by the ordinary spectral triple $(\mathcal{A}_{\mathfrak{J}}, D_{\mathfrak{J}}, \mathcal{H}_{\mathfrak{J}})$ corresponding to $(\mathcal{A}, D, \mathcal{H})$ via the "Wick rotation" induced by an \emph{admissible}\footnote{Compare \cite{Stro}}     fundamental symmetry $\mathfrak{J}$: \emph{i. e.}  we propose the Dirac operator $D$ to be so chosen that the index map  induced by $D_{\mathfrak{J}}$  on $\mathcal{A}$ coincides with the index map in the construction of the necessary conditions. However this topological-type condition embracing only the global aspect of the theory may be insufficient for reconstruction of $D$ (even in this undeformed, i. e.  free-field case). One may hope to reconstruct in this way only the sign\footnote{Of course modulo a trivial modification on the kernel, but preserving the index, so that index $F$ = index $D_{\mathfrak{J}}$ , compare \emph{e. g.} \cite{Connes}.}     $F = D_{\mathfrak{J}}|D_{\mathfrak{J}}|^{-1}$ of $D_{\mathfrak{J}}$.\footnote{Independently of this many examples of Fredholm modules -- bounded versions of ordinary riemannian spectral triples, connected to free (quantum) fields has been constructed, at least for fields without any gauge freedom. Compare e. g. \cite{GarVar}, where it is shown how the free fermion charged fields give rise to natural constructions of Fredholm modules. In the same book \cite{GarVar}, Chap. IV.13, connection of the adiabatic limit and the Bogoliubov-Epstein-Glaser local renormalization with the local index formula is noticed and emphasized.}        We expect, however, that the full reconstruction of $D$ in the undeformed (i. e. free field) case will be difficult -- we indicate a method of reconstruction of undeformed $(\mathcal{A}, D, \mathcal{H})$ on Minkowski spacetime in Sect. ~\ref{freeD}. The local information which shows up in the microlocal renormalization is useless for the reconstruction of "undeformed" $D$.  But if the undeformed $D$ was unknown, then any effort to proceed the other way round after Fedosov and investigate the \emph{sufficiency condition} for the existence of the asymptotic representation would be hopeless (still in QED). It shoud be stressed that already in solving the second task we will need to know the undeformed 
$(\mathcal{A}, D, \mathcal{H})$ on minkowski background in order to reformulate the asymptotic conditions of Blanchard and Seneor in terms of symbol calculus -- the immediate analogue of the asymptotic properties of the Weyl representation
on $\mathbb{R}^{2n}$. We propose to make only first steps towards this goal. We assume that we have undeformed ordinary (riemannian) spectral tripe 
$(\mathcal{A}, D_{\mathfrak{J}}, \mathcal{H}_{\mathfrak{J}})$ and that it is compact ( $\mathcal{A}$ unital). Now we could incorporate the microlocal renormalization of Brunetti and Fredenhagen \cite{BruFred} to utilize the local information.

\vspace*{0.5cm}

\begin{tabular}{|p{11.2cm}|}\hline
Thus we have arrived at the third task of our proposal: to construct a formal deformation $(\tilde{\mathcal{A}_{\mathfrak{J}}}, \tilde{D_{\mathfrak{J}}}, \tilde{\mathcal{H}_{\mathfrak{J}})}$ of $(\mathcal{A}_{\mathfrak{J}}, D_{\mathfrak{J}}, \mathcal{H}_{\mathfrak{J}})$ along the lines of D\"utsch and Fredenhagen (or Bordeman and Waldmann), thus to investigate stability of the spectral triple structure $(\mathcal{A}, D_{\mathfrak{J}}, \mathcal{H}_{\mathfrak{J}})$ under deformation, \emph{i. e.} try to prove the analog of the above stability theorem for compact spectral triple. \\ \hline
\end{tabular}

\vspace*{0.5cm} 

If the stability is preserved, then we can expect to have the full analog of the Fedosov theorem (in compact case only) and imitate the main steps of Fedosov having the full abstract calculus of symbols worked out by Connes and Moscovici \cite{ConMos} for the undeformed $(\mathcal{A}, D_{\mathfrak{J}}, \mathcal{H}_{\mathfrak{J}})$. Again we expect that even in this simplified case (QED: existence of Green functions assured) the full analog of Fedosov theorem will be difficult to work out, as the non-compact triples involve much more technicalities\footnote{In fact Fedosov proved his theorem on sufficiency for the existence of asymptotic representation for compact manifolds only. But an analogue theorem is certainly true for the non-compact case as well (after some reasonable assumptions of course).}. Yet the full version (\emph{necessity} and \emph{sufficiency}) would be very desirable as we expect in this case that the integrality of the index (\emph{necessity} and \emph{sufficiency} condition) puts strong limitations on the allowable values of the deformation parameter, 
\emph{i. e.} on the coupling constant $g$. This goes outside our proposal, but we expect that in general situation (not only for QED) an analog of Fedosov theorem holds: namely that the actual asymptotic representation does exist (and so the adiabatic limit exist) whenever the index induced by $(\mathcal{A}, D, \mathcal{H})$\footnote{More exactly: by the corresponding $(\mathcal{A}, D_{\mathfrak{J}}, \mathcal{H}_{\mathfrak{J}})$.}     fulfills some integrality conditions. By what we already know of the charge structure from the algebraic quantum field theory\footnote{Compare the Doplicher, Haag and Roberts analysis in ref. \cite{Haag}.}        we expect that such index describes charge structure of the theory.  Because on the other hand the properties of the index reflect universal properties of the (non-commutative but \emph{classical}) spacetime, the charge structure\footnote{As far as reflected  by the index.}   would come out of (non-commutative) spacetime properties. At this place I quote a problem posed by Staruszkiewicz \cite{Star1}:  \emph{How is it  possible at all that the electric charges in general, and the electric charges of particles so much diverse as leptons and hadrons in particular, are all equal to the multiples of one and the same universal unit charge? How is it possible that the electric charge of electron and the electric charge of proton are equal with an an unusually small experimental error, such that their ratio is equal to $1$ with  the experimental error less than $10^{-21}$}?  We agree with A. Staruszkiewicz that the simplest explanation of this problem is to assume that the electric charges of proton and electron are mathematically equal and that the charge structure (in particular the property of the electric charge cited above) reflects a property of spacetime and not properties of particles themselves, just as for spin, whose properties reflect the rotation symmetries -- a subgroup of spacetime symmetries, and result from the properties of irreducible unitary representations of the subgroup. The problem of Staruszkiewicz is an important motivation for this proposal. However the hypothesis presented here differs significantly from the theory proposed by Staruszkiewicz \cite{Star2}. Here we intend to reconcile the puzzle of Staruszkiewicz with the observed fact that the electric charge (and generalized charges, such as baryon number, lepton number, generalized isospin, ...) do not superpose similarly as macroscopic immediately observable quantities, and propose a tentative hypotheses that the generalized charges do not superpose. Staruszkiewicz adopts the initial assumption that the electric charge can in principle at least superpose\footnote{This is rather an artifact of the (possibly an oversimplifying) assumption that the regime of validity of the ordinary quantum mechanics is unrestricted then of the Staruszkiewicz's theory itself (one has to assume at least that the Coulomb field falls within the regime). In this approach an \emph{ad hoc "vector reduction mechanism"} is needed.}, and consequently, that the phase of wave function -- a degree of freedom canonically conjugate to the charge emerging from the U(1)-gauge, is subject to quantization. Thus he lives open the question: why we do not observe any coherent superpositions of electric charges? These assumptions (of this proposal and that of Staruszkiewicz) lead to different conceptions of spacetime. 

	What are the conceptual and computational gains of the hypothesis proposed here and of  the conception of non-commutative spacetime adopted here? Perhaps it is worth to emphasize that the inclusion of the algebra of spacetime coordinates as a structural ingredient of the theory along the lines proposed here allows in principle to keep the particle interpretation on curved spacetime, even if the spacetime does not posses any time-like Killing vector fields, considering the relationship between the algebra of spacetime coordinates and the algebra of detectors. This allows (potentially) to make a practical use of the renormalization theory of Brunetti and Fredenhagen. Indeed we can, in principle at least, pick up the vacuum-like states by incorporating the relationship between annihilators and detectors. This would give a solution to the well known problem set for e. g. by Buchholz in section 8 of his review article \cite{Buch}. We should emphasize that the geometric method proposed here introduces a whole variety of non-commutative geometry tools and connects them with the existence problem for the adiabatic limit, a problem which is still open (to the author's knowledge) for theories with non-abelian gauge symmetry (\emph{confinement}). Last but not least: we get for free the time's arrow for non-superposing quantities, as a consequence of the non-commutative character of the algebra of spacetime coordinates. It is expected of any 'reliable' theory embracing macroscopic non-superposing quantities not only by theorists like Haag or Penrose, but most of all by those theoreticians who have everyday contact with quantum chemical and optical laboratory, see e.g. \cite{Prim} Compare also the "complementarity concept" of Bohr \cite{Bohr}.

\section{Undeformed $(\mathcal{A}, D, \mathcal{H})$}\label{freeD}

As we have mentioned above to give the asymptotic conditions of Blanchard and Seneor the shape of asymptotic conditions of Fedosov (for the asymptotic operator representation) i.e. in terms of asymptotic conditions imposed on the symbols of operators, we need two things: 1) to pass from Green functions to operators (after Wightman); 2) then wee need ``undeformed'' $(\mathcal{A}, D, \mathcal{H})$ on Minkowski spacetime, i.e.  for free QED on Minkowski background (the immediate analogue of the Weyl representation $W_{D}(\mathbb{R}^{2n})$ on the symplectic manifold $\mathbb{R}^{2n}$). To this end the knowledge of the "undeformed" $(\mathcal{A}, D, \mathcal{H})$ (for free QED on Minkowski background) is sufficient provided the spectral triple $(\mathcal{A}, D, \mathcal{H})$ is stable under deformation. 

Here we concentrate on the most difficult part of our proposal, and indicate a way 
of constructing undeformed $(\mathcal{A}, D, \mathcal{H})$. Here we restrict attention
to the case based on Minkowski background, i.e. for free QED on Minkowski background,
as we are primarily interested in reformulating the asymptotic conditions of Blanchard and Seneor in geometric terms explicitly involving the triple 
$(\mathcal{A}, D, \mathcal{H})$ and the abstract symbol calculus.

From general principles of QFT and especially from the experimental
verification of the celebrated ``dispersion relation'' we expect the spacetime 
algebra $\mathcal{A}$, or rather its geometric structure encoded in 
$(\mathcal{A}, D, \mathcal{H})$ to be ``fairly classical''. What is important 
here is to give it the spectral operator format $(\mathcal{A}, D, \mathcal{H})$
allowing to a noncommutative deformation. 
In particular one can expect $\mathcal{A}$ to be already
(Morita equivalent to) a commutative algebra in the free field case
and try to identify it within the field algebra together with the Dirac
operator $D$ in the free field Krein space $\mathcal{H}$.\footnote{However
we cannot expect a priori that the algebra $\mathcal{A}$ is cyclic as 
represented in $\mathcal{H}$, and if every element of $[D,\mathcal{A}]$ 
preserves the cyclic subspaces, allowing to reducibility of 
the spectral triple $(\mathcal{A}, D, \mathcal{H})$ and thus several 
(infinite and discrete or continuous) sum of disjoint connected 
components of sp$\mathcal{A}$ (or so to say: discrete sum  or "continuous sum" of classical spacetime copies) with high multiplicity.} 

\vspace{0.5cm}

{\bf Remark 1}  It should be stressed that the microlocal renormalization of 
Brunetti and Fredenhagen works well also for general curved globally hyperbolic spacetimes. In fact we expect that topology of $(\mathcal{A}, D, \mathcal{H})$ for free QED on Minkowski spacetime trivializes (we expect that we recover spectrally exactly the ordinary spacetime geometry). It is in the general case with curved spacetime with non-trivial topology where we expect that the index type conditions
will show up. Of course to realize our proposal in QED case, we will need to construct
the undeformed $(\mathcal{A}, D, \mathcal{H})$ in every case where the deformation works, i.e. for general globally hyperbolic spacetimes. But this is of course unnecessary for the geometric Fedosov-type formulation of Blanchard-Seneor asymptotic conditions on Minkowski spacetime.

{\bf Remark 2} One can argue perhaps that the proposed method gives only a (rather sophisticated) geometric form to the Blanchard-Seneor theorem about Green functions in QED capable of investigation of the adiabatic limit for QED in the Brunetti-Fredenhagen
renormalization on curved globally hyperbolic spacetimes. And argue further that without the respective analogue of Blanchard-Seneor for gauge field theories with confinement 
it will give us nothing towards confinement. But this opinion would be premature for at least one reason. The proposed geometric reformulation of what we essentially know about asymptotics in QED on Minkowski spacetime has the important property that it depends on the undeformed triple $(\mathcal{A}, D, \mathcal{H})$, and this triple in turn depends on the free theory in question which has an immediate influence upon the construction of symbols in the asymptotic conditions formulated geometrically after Fedosov. We can not exclude at this stage before our proposal is completed just for QED, that the replacement of $(\mathcal{A}, D, \mathcal{H})$ (with that corresponding to respective free field(s) with confinement) in the symbol calculus of geometric asymptotic conditions will do work. That the Fedosov theorem is empty in case of flat symplectic manifold $\mathbb{R}^{2n}$, as the topology of $\mathbb{R}^{2n}$ is trivial, and all deformation parameters within the interval $[0,1)$ are allowed (just as we expect in our case for Minkowski spacetime, with the weakest restriction on the coupling constant(s)) is completely unimportant here.

\vspace{0.5cm}

Construction of undeformed $(\mathcal{A}, D, \mathcal{H})$
on Minkowski background proposed here is reduced to the construction
of a "Fourier transform" on a uniform (pseudo)riemannian  
manifold acted on by the Poincar\'e group.
This is suggested by the following three sources: 1) free field construction for 
particles with strictly positive mass, or better for fields constructed out of these irreducible (unitary) representations of the Poincar\'e group which have strictly positive mass operator; 2) by the Haag-Ruelle formulation of scattering theory for QFT
with the vacuum strictly separated from all other states by a mass gap; and 3)
by the nonrelativistic quantum field theory.
Indeed the construction of free field (out of the irreducible representations of the Poincar\'e group with strictly positive mass) as well as the construction of
one-particle states in Haag-Ruelle theory (with
a positive lower bound in the spectrum of the mass operator in the subspace orthogonal
to the vacuum) is strictly analogous to the construction of an "inverse Fourier transform" relating the spectrum of momentum operators (translation generators) 
with the spectrum of the Schwartz algebra $\mathcal{S}(\mathcal{M})$ on the 
Minkowski manifold $\mathcal{M}$, i.e. with spacetime points. 

In order to explain this we remind the rudiments of harmonic analysis on uniform manifolds. Suppose we have a uniform differential riemannian (or pseudoriemannian) manifold $\mathcal{M}$ of dimension $n$ (in fact we consider also manifolds $\mathcal{M}$ with more degenerate geometric structure, such as e.g. the Galilean spacetime with the Galilean group acting on it) acted on by a Lie group $G$, with a (pseudo)metric form $g$ invariant under $G$. Then we consider the
Hilbert space\footnote{In fact we intent here a more general case of Krein-type space, but for a while the reader may initially assume $\mathcal{H}$ to be ordinary Hilbert space.} $\mathcal{H} = L^2(\mathcal{M}, d\upsilon)$ of square integrable functions 
with respect to the invariant volume form $d\upsilon$ (as is standard in works of Gelfand, Harish-Chandra and others on harmonic analysis), however we will be primarily interested in Hilbert spaces $\mathcal{H}$ (or Krein spaces) of square integrable spinors or sections of more general Clifford modules over $T^*\mathcal{M}$, although this is unimportant in presenting the general idea. One then consider the unitary regular right representation $T$ of $G$ acting in $\mathcal{H}$ and an appropriate algebra $\mathcal{S}(\mathcal{M})$ of functions of fast decrease with nuclear\footnote{Wed need nuclearity to construct generalized proper vectors (or explicit decompositions into continuous sums/integrals) as weak derivations of vector valued spectral measures of the appropriate differential selfadjoint operators invariant under $G$, and thus commuting with $T$, compare \cite{Gelfand}.} Fr\'echet space as a linear topological space (just the algebra of smooth functions in case of compact $\mathcal{M}$). We can consider also the algebra $\mathcal{S}(\mathcal{M})$ as acting in 
$\mathcal{H}$ as a multiplication algebra with point wise multiplication. The regular representation $T_g$ induces the transformation $a \mapsto T_gaT_g^{-1}$ of
$a \in \mathcal{S}(\mathcal{M})$ coinciding with the ordinary group action $T_gaT_g^{-1}(x) = a(g^{-1}x)$ for functions on $\mathcal{M}$. Harmonic analysis ("Fourier transform" on $\mathcal{M}$) corresponds to a decomposition of the regular right representation $T$ acting in $\mathcal{H}$ into continuous sum (integral) of irreducible sub representations. To this decomposition there corresponds a decomposition of every element $f$ of $\mathcal{H}$ into continuous sum (integral) of its components belonging to the irreducible generalized subspaces -- the "inverse Fourier integral of $f$". For example the (inverse) Gelfand Fourier transform on Lobachevsky space (acted on by $SL(2,\mathbb{C})$) together with the respective algebra $\mathcal{S}(\mathcal{M})$ has been constructed in \cite{Gelfand}.\footnote{Gelfand and his co-workers \cite{Gelfand} consider regular representations acting on square integrable functions. One can do the same of course for regular representations acting e.g. on spinors. For example every known spectral triple explicitly constructed on a manifold uniform for a Lie group has been constructed with an implicit or explicit help of harmonic analysis.} Now it is important that in general such situation one can construct decomposition of elements $f$ of $\mathcal{H}$ (inverse Fourier transform) in purely spectral manner. We consider a maximal commutative algebra $\hat{\mathcal{A}}$ generated by representors of generators of one parameter subgroups (or their functions). Let $\hat{\mathcal{A}}$ be generated by $P_1, \ldots, P_n$ and consider their joint spectrum sp$(P_1, \ldots, P_n)$ (in particular for the Euclidean n-space $\mathbb{R}^n$ these could be chosen to be translation generators along the canonical coordinates' in case of $SL(2, \mathbb{R})$ 
acting on the Lobachevsky plane $\mathcal{L}^2$ we may chose $P_1$ to be the Casimir operator, i.e. the Laplacian on the Lobachevsky plane, and for $P_2$ we may chose a generator of a one-parameter boost subgroup; in our case we will consider translation generators). Then we will have

\begin{equation}\label{dec}
f(x) = \int \limits_{\textrm{sp$(P_1, \ldots, P_n)$}} \mathcal{F}f(s) \Theta(x;s) d\nu(s) ; \,\,\,  \mathcal{F}f(s) = \int \limits_{\mathcal{M}} f(x) \Theta(x;s) d\upsilon(x)  
\end{equation}          
where $\Theta(x;s)$ is a complete set of common generalized proper functions of the operators $P_1, \ldots, P_n$ corresponding to the joint spectral point $s$
of sp$P_1, \ldots, P_n$, and $d\nu(s)$ is their joint spectral measure. In fact the Fourier transform on the Lobachevsky plane or space in \cite{Gelfand} has not this clear spectral form as no generators (besides the Laplacian) $P_1, \ldots, P_n$ of the algebra of operators, which are simultaneously diagonalized by the Fourier transform  are explicitly constructed.\footnote{However one can easily modify their Fourier transform to obtain one diagonalizing generator of rotation and the Laplacian.} Existence of Fourier transforms diagonalizing say the Laplacian on $\mathcal{L}^2$ and generator of one of boost subgroups follows from general theory presented in \cite{GelfandIV}, \cite{Gelfand}, and by the same theory existence  of Fourier transforms
is assured, diagonalizing any maximal algebra $\hat{\mathcal{A}}$ of functions of generators of the regular representation $T$. (In a subsequent paper we give an explicit construction of the Fourier transform on $\mathcal{L}^2$ diagonalizing the Laplacian and a generator of boosts.) Of course from the continuous sum (integral) decomposition one can recover the decomposition into irreducible generalized subspaces, in particular in the case of the classical groups considered here, we decompose the joint spectrum manifold sp$(P_1, \ldots, P_n)$ into sub manifolds sub($\lambda$) of constant values 
$\lambda$ of the Casimir operator of the whole Group. Then the integral 
\begin{equation}\label{decom}
f_\lambda(x) = \int \limits_{\textrm{sub($\lambda$)}} \mathcal{F}f(s) \Theta(x;s) d\nu_\lambda(s) ; \,\,\,  \mathcal{F}f(s) = \int \limits_{\mathcal{M}} f(x) \Theta(x;s) d\upsilon(x)  
\end{equation}
 over the sub manifold sub($\lambda$) with the measure $d\nu_\lambda(s)$ induced 
by $d\nu(s)$ gives the generalized invariant subspace corresponding to the proper value $\lambda$ of the Casimir operator $C$. Now we may write the decomposition (\ref{dec})
in the following form 
\begin{equation}\label{decom}
f(x) = \int \limits_{\textrm{sp$C$}} f_\lambda (x) d\lambda ; \,\,\, f_\lambda(x) = \int \limits_{\textrm{sub($\lambda$)}} \mathcal{F}f(s) \Theta(x;s) d\nu_\lambda(s) 
\end{equation}
obtaining the decomposition of the Hilbert space $\mathcal{H}$ as a continuous
sum/integral 
\[
\mathcal{H} = \int \limits_{\textrm{sp$C$}} \mathcal{H}_\lambda d\lambda
\]
of irreducible subspaces $\mathcal{H}_\lambda$ of generalized functions 
of the form 
\[
f_\lambda(x) = \int \limits_{\textrm{sub($\lambda$)}} \mathcal{F}f(s) \Theta(x;s) d\nu_\lambda(s); \,\,\, f \in L^2(\mathcal{M}, d\upsilon). 
\]
Thus the ``Fourier transform'' provides a unitary transformation diagonalizing 
the operators of the algebra $\hat{\mathcal{A}}$ and the ``inverse Fourier transform'' diagonalize the algebra $\mathcal{S}(\mathcal{M})$ viewed as multiplication algebra 
in $\mathcal{H}$. In this sense the algebras $\mathcal{S}(\mathcal{M})$ and 
$\hat{\mathcal{A}}$ are dual to each other.   

In general the the manifold sp$(P_1, \ldots, P_n)$ is discrete sum of connected 
components, and thus have a mixed character: continuous and discrete, depending on the choice of the generators $P_i$. However it is purely discrete only for compact
$\mathcal{M}$. In general in decomposition of an element $f \in L^2(\mathcal{M}, d\upsilon)$ there participate all irreducible subspaces $\mathcal{H}_\lambda$ of all irreducible sub representations of $T$. The same will be true for the Minkowski spacetime
$\mathcal{M}$ acted on by the Poincar\'e group represented in the Clifford bundle 
used in \cite{Stro} of square integrable sections (with the algebra 
$C^{\infty}(\mathcal{M})$ replaced with $\mathcal{S}(\mathcal{M})$). 

However in some situations (we will give physically important examples below) 
the full harmonic analysis on the whole spacetime manifold $\mathcal{M}$ involving all irreducible sub representations of $T$ is unnecessary in recovering the dual relationship between the spectra of $\mathcal{S}(\mathcal{M})$ and $\hat{\mathcal{A}}$. 
In particular it may happen that after restricting the integration in (\ref{dec})
over sp$(P_1, \ldots, P_n)$ to an integration over a fixed sub manifold, say to the 
sub manifold sub($\lambda$), corresponding to the irreducible subspace
$\mathcal{H}_\lambda$ (irreducible sub representation 
$T_\lambda$ of $T$), and after restricting the argument $x$ in the integrand
of (\ref{dec}) to a sub manifold sub($\mu$) of $\mathcal{M}$, one obtains
a Fourier transform i.e. a unitary map between 
$L^2(\textrm{sub($\mu$)}, d\upsilon_{\textrm{sub($\mu$)}})$
and $L^2(\textrm{sub($\lambda$)}, d\nu_{\textrm{sub($\lambda$)}})$.
Besides the sub manifolds sub$(\mu)$ of $\mathcal{M}$ compose a one-parameter foliation (codimension one foliation) of $\mathcal{M}$, parametrized by the real number $\mu$. This is a rather very exceptional situation, strongly depending on the group structure of the group $G$ in question and on the uniform manifold $\mathcal{M}$ acted on by $G$. It may even happen that although one has to use a Krein-type space $\mathcal{H}$ in order to encode the algebra $\mathcal{S}(\mathcal{M})$ together with the metric structure
of $\mathcal{M}$ spectrally in the Connes-Strohmaier format the irreducible 
subspace $\mathcal{H}_\lambda$ degenerates to an ordinary Hilbert space. Of course 
in the restricted integral 
\[
f_\lambda(x) = \int \limits_{\textrm{sub($\lambda$)}} \mathcal{F}f(s) \Theta(x;s) d\nu_\lambda(s) 
\] 
one can consider the function (cross section
of the respective Clifford module) $f$ with the argument ranging all over
the manifold $\mathcal{M}$ not restricted to any of the sub manifolds
sub($\mu$) with fixed value of $\mu$. But then the function/cross section $f$
will not be square integrable and will not belong to $\mathcal{H}$ but will fulfil
a differential ``wave equation'' corresponding to the irreducible sub representation 
$T_\lambda$ of $T$. Of course this restricted Fourier transform works for all sub manifolds sub($\mu$) of $\mathcal{M}$ for each $\mu$ separately. In recovering
$f \in \mathcal{H}$ on the whole spacetime $\mathcal{M}$, as in (\ref{dec}), 
using just one irreducible subspace $\mathcal{H}_\lambda$ (for fixed $\lambda$) is of course insufficient and in general all the irreducible subspaces $\mathcal{H}_\lambda$ will participate in the decomposition (\ref{dec}). Such exceptional situation
allowing to a construction of ``restricted Fourier transform'' we encounter in case
of Bargmann central extension\footnote{Although action of the extension on the Galilean spacetime degenerates to the ordinary action of the inhomogeneous Galilean group, using of the Bargmann extension is essential if one intents to describe spectrally the Galilean spacetime manifold, see below for some further comments.} $G$ of inhomogeneous Galilean group acting on the Galilean spacetime $\mathcal{M}$. The sub manifold 
sub($\lambda$) corresponds to the paraboloid of constant mass equal 
$\lambda$ in the four-momentum space\footnote{Where the generator of $P_4 = M$ of the center of the Bargmann extension has to be added to the translation generators 
$P_0, \ldots, P_3$, which in quantum mechanics is physically interpreted as mass operator.}  
sp$(P_0, P_1 \ldots, P_3, P_4 = M)$ and the sub manifolds sub($\mu$) of $\mathcal{M}$ correspond to to the simultaneity hyperplanes $t = \mu$. In fact if we wants to describe the Galilean spacetime spectrally and explain in addition its connection to 
non-relativistic quantum fields, it is the central Bargmann 
extension\footnote{This extension 
$G$ may be realized e.g. as a product $ G' \times \mathbb{R}$, where $G'$ is the inhomogeneous Galilean group with the following multiplication rule: $(\theta,r)\cdot(\theta',r')
= (\theta + \theta' + \xi(r,r'), rr')$, where $\xi$ is the standard Bargmann exponent
on the inhomogeneous Galilean group $G'$ equal to 
\[
\xi(r,r') = \frac{1}{2}\big[ {\bf u}\cdot R {\bf v}' - {\bf v} \cdot R{\bf u}' 
+ \eta' {\bf v} \cdot R {\bf v}'  \big],
\]
(dot $\cdot$ stands for the ordinary scalar product and $R$ denotes the rotation
matrix) for the inhomogeneous Galilean transformations $r$ and $r'$:
\[
r: ({\bf x}, t) \mapsto (R{\bf x} + t{\bf v} + {\bf u}, t + \eta) \,\,\, \textrm{and} 
\,\,\, r': ({\bf x}, t) \mapsto (R'{\bf x} + t{\bf v}' + {\bf u}', t + \eta').
\]
Note that $Z = \{(1, \theta), \theta \in \mathbb{R} \}$ is a central subgroup of the extension $G$, as $\xi(1,1) = \xi(1,r) = \xi(r,1) = 0$, $r \in G'$, but $\{(r,0), r \in G'\}$ is not a subgroup. However $G/Z$ is group isomorphic to the inhomogeneous Galilean group $G'$. Commutation rules for the generators of the Bargmann extension $G$ have the following form:
\begin{multline}
[J_i, J_j] = i\epsilon_{ijk} J_k, \,\,\, [J_i,K_j] = i \epsilon_{ijk} K_k \,\,\,
[J_i, P_j] = i \epsilon_{ijk} P_k \\
[K_i, P_0] = i P_i  \,\,\, [K_i, P_j] = i \delta_{ij} M \,\, (= i\delta_{ij} P_4) \\
[J_i, P_0] = [K_i, K_j] = [P_i,P_j] = [P_j, P_0] 
= [J_i, M] = [K_i, M] = [P_i, M] = [P_0, M] = 0.
\end{multline}
where $J_i, K_i, P_i$ are generators of rotations, proper Galilei transformations, space translations, and $P_0, M$ are generators of time translations and of the central subgroup $Z$ respectively.}
$G$ of the inhomogeneous Galilean group which is more natural here as a symmetry group. Indeed the appropriate Dirac operator\footnote{Although for general Dirac-type operator on $\mathcal{M}$ with a pseudo-riemannian metric, or even with a more degenerate ``metric'' structure, there does not exist any natural Hilbert space acted on by $D$, such that $D$ is (essentially) selfadjoint, there nevertheless does exist natural Krein-type space with respect to which the operator $D$ is selfadjoint in the generalized Krein sense, compare e.g. \cite{baum}.} we should use here is the non relativistic Dirac operator $-i\partial_t \otimes A - i\partial_i \otimes B^i + 1 \otimes C$ found by 
L\'evy-Leblond \cite{levy1}, where $A, B^i, C$ are elements of a Clifford algebra over the (five dimensional) extension\footnote{Corresponding to the Bargmann extension of the Galilean group.}  of tangent space with a positive definite and singular quadratic form in it. Indeed, in this Galilean case the the Krein-type space $\mathcal{H}$ corresponding to the more degenerate ``metric'' structure of 
$\mathcal{M}$ is slightly different from the ordinary Krein space and may be reduced to a positive definite inner product space with non-trivial closed subspace of zero norm vectors, which cannot be quotiened out, as the kernel subspace reflects the degenerate character of the ``metric'' structure on 
$\mathcal{M}$. This can be immediately seen from the results of \cite{levy1}, as all the algebra elements $A, B^i, C$ can be (algebraically) generated from the standard Clifford algebra over a positive definite metric of rank 4. Indeed as follows from \cite{levy1}, we may put: $A = -i/2(\beta +\beta \gamma^4), C = mi(\beta - \beta \gamma^4), B^i = \beta \gamma^i$, where $\gamma^\alpha$, $1 \leq \alpha \leq 4$ are the standard generators of the Clifford algebra over four dimensional vector space with euclidean metric form: $\gamma^\alpha \gamma^\rho + \gamma^\rho \gamma^\alpha = 2 \delta^{\alpha \rho}$, $1 \leq \alpha, \rho \leq 4$; and where $\beta$ is an arbitrary non-singular $4 \times 4$ matrix, which therefore may be chosen to be equal say $\gamma^4$. The action of the Bargmann extension $G$ in the (degenerate) space $\mathcal{H}$ of square integrable non relativistic bispinors of L\'evy-Leblond as well as the invariance asserting that his Dirac operator commutes with this action immediately follows from \cite{levy1}.\footnote{We do not go into details here as we are not primary interested in Galilean spacetime.} Let us give more involved and non-trivial physically interesting examples pertaining to quantum field theory.   
 
\vspace{0.5cm}

{\bf Example 1} Consider an irreducible unitary representation $U_{m,\alpha}$ of the Bargmann central extension $G$ of the inhomogeneous Galilean group, acting in an ordinary Hilbert space $\mathcal{H}_{m, \alpha}$, corresponding to the mass $m$ and spin $\alpha$. Then using $U_{m,\alpha}$ let us construct in the standard way a free non relativistic quantum field acting in the Fock space
\[
\mathcal{H}_{F,1}  = \left\{ \begin{array}{ll}
\oplus_{N = 0}^{\infty} (\mathcal{H}_{m, \alpha}^{\oplus N})_S & \textrm{for bosons}\\
\oplus_{N = 0}^{\infty} (\mathcal{H}_{m, \alpha}^{\oplus N})_A & \textrm{for fermions}
\end{array} \right.
\]
together with the unitary representation
\[
U_1 = \left\{ \begin{array}{ll}
\oplus_{N=0}^{\infty}\big( U_{m, \alpha}^{\otimes N} \big)_{S} & \textrm{for bosons}\\
\oplus_{N=0}^{\infty} \big( U_{m, \alpha}^{\otimes N} \big)_{A} & \textrm{for fermions}
\end{array} \right.
\]
of $G$ acting in $\mathcal{H}_{F,1}$. In the above formulas the $N=0$ summand equals just
$\mathbb{C}$ with the natural inner product in $\mathbb{C}$ and represents states proportional to the vacuum, with the trivial representation of $G$ in $\mathbb{C}$ (every element of $G$ is just represented as multiplication by 1). 
Now to this free field we add a free ``Galilean
electromagnetic field'', which from the quantum field theory point of view has 
the character of a pure ``admixture'', which normally could be completely ignored, 
but it is crucial in  connection to the
harmonic analysis on the whole spacetime $\mathcal{M}$. 
Namely we consider in addition  a ``free Galilean electromagnetic quantum field'', 
which is composed of the zero mass irreducible representations of the Bargmann extension $G$ of the inhomogeneous Galilean group, just like the ordinary quantum field is composed out of the massive irreducible representations of the Bargmann central extension of the inhomogeneous Galilean group. Then we compose together a free system of Galilean uncoupled fields, the
former particle field together with the ``Galilean electromagnetic 
field''.  Because in the standard reference articles, e.g. \cite{levy1},  treating Galilei invariant description of quantum particles (irreducible representations of the inhomogeneous Galilean group) zero mass helicity 1 four-vector particles are not analysed from the quantum mechanical point of view and because in the known works on Galilean quantum field theory, e.g. \cite{levy2}, physical triviality/non triviality of Galilei invariant quantum electromagnetism is not critically treated
from the group representation point of view, we indicate here the basic idea. Namely we analyse the unitary irreducible representations of the group\footnote{Here we intend to analyse zero mass representations, so that that the representation of the Bargmann extension $G$ degenerates to an irreducible representation of the ordinary inhomogeneous Galilean group $G'$ in the zero mass irreducible subspace. The group extension $0 \mapsto Z \mapsto G \mapsto G' \mapsto 1$ becomes trivial there, as the representation of induced on the central subgroup degenerates to just identity operator. For non zero mass case the representation of Bargmann extension is faithful and the irreducible representation of $G$ does not degenerates to a unitary representation of $G'$. For the non-zero mass case the irreducible unitary representations of the inhomogeneous Galilean group $G'$ are not interesting as the one-particle states composed of them lack any reasonable locality properties -- no reasonable position operator exists for them.} $G$ acting in $\mathcal{H}$ exactly as Wigner did for the Poincar\'e group decomposing the space $\mathcal{H}$ with respect to (the spectrum of) the commuting generators of translations and the generator $P_4 = M$ of the center $Z$.

In should be stressed that here (and it lies at the very heart of the construction of one-particle wave functions in QFT) some additional assumptions not entirely controlled are implicitly used at this place. Namely we have to assume that we have a sufficiently well behaved unitary representation of $G$ in a Hilbert space $\mathcal{H}$, treated as if it was a regular representation acting on the spacetime manifold 
$\mathcal{M}$, so that the vector-valued spectral measure of the generators $P_0, \ldots P_3, P_4$ (composing  a maximal commuting set of generators of the group $G$ in question) can be weakly differentiated and by this could give us a ``Fourier-type'' construction of one-particle wave functions. On the other hand we have mathematically well defined construction of free fields and the mentioned assumptions, not entirely clear, must be reflected somehow by the mathematical structure of free fields. This is in fact our immediate task in extracting them more accurately, as the extraction involves spectral reconstruction of spacetime manifold out of free fields. 

Let us proceed for a while after Wigner in analyzing the representation of $G$. As we have said we decompose $\mathcal{H}$ with respect to $P_0, \ldots, P_3. P_4$ (energy, momentum and mass operators)
\[
\mathcal{H} = \int \limits_{\textrm{sp$(P_0, \ldots, P_3, P_4)$}} \mathcal{H}_pd\nu(p)
\]
into a direct integral of generalized common proper subspaces of
$P_0, \ldots, P_3, P_4$. After restricting the continuous integral decomposition to the paraboloid $p_1^2 + \ldots p_3^2 = 2\lambda p_0^2$ of mass $m = \lambda$, i.e. to a sub manifold sub($\lambda$) of sp$(P_0, \ldots, P_3. P_4)$, and using the Wigner's technique of ``little Hilbert space'' we obtain the irreducible subspace $\mathcal{H}_\lambda$  corresponding to an irreducible sub representation in the following form 
\[
\mathcal{H}_\lambda 
= \mathfrak{h} \otimes \mathfrak{H}, \,\,\, \textrm{where} \,\,\, 
\mathfrak{H} = L^2(\textrm{sub($\lambda$}), d\nu_\lambda(p)),
\]
with respect to the measure $d\nu_\lambda(p)$ on sub($\lambda$) induced by the spectral measure $d\nu(p)$; and $\mathfrak{h}$ is the ``little Hilbert space''. Thus the elements of the irreducible subspace $\mathcal{H}_\lambda$ are $\mathfrak{h}$-valued functions
$p \mapsto \hat \Psi(p)$ on sub($\lambda$) of the momentum $p$. Then we proceed just like {\L}opusza\'nski did \cite{lop1, lop2} for the Poincar\'e group, in showing that the inverse Fourier transform  ''restricted'' to the sub manifold sub($\lambda =0$), i.e. in case of zero mass $m = \lambda = 0$: 
\begin{multline}\label{0m-gal}
\Psi(x) = (2\pi)^{-3/2} \int 
\limits_{\textrm{sub($\lambda$)}} \hat \Psi(p) e^{i({\bf px} - p_0 t)} d\nu_\lambda(p)\\
= (2\pi)^{-3/2} \int \hat \Psi({p_0, 0,0,0}) e^{-ip_0 t} dp_0, \\ \textrm{where} \,\,\, 2 \lambda p_0 = {\bf p}^2 \,\,\, \textrm{so} \,\,\, {\bf p} = 0,
\end{multline}
is a four-vector field on spacetime $\mathcal{M}$ with helicity 1 if and only if the (irreducible) representation space $\mathcal{H}_\lambda$ is a completely degenerate inner product space with the inner product identically equal zero; or more precisely if and only if the ''little space'' $\mathfrak{h}$ is not ordinary Hilbert space, but a finite dimensional space with completely
singular inner product, i.e. with all vectors having the zero norm. In fact here for the group $G$ the ``little'' group is equal to the Euclidean group $E_3$ and not $E_2$ as for the (double cover of the) Poincar\'e group and the situation is slightly more involved for analysis then for the Poincar\'e group. Let us call this representation  
$U_{0,1}$. This is no surprise as the Hilbert space structure has always to be modified into a Krein-type (or more degenerate) space if one uses single particle wave functions with redundant components, e.g. in describing spin 1 particle by a four-vector wave functions. Our case is even more degenerate, as we expect no real free quantum particles in the Galilei invariant quantum electromagnetic field.\footnote{Recall that according to \cite{lebel} there are two different non relativistic limits of classical Maxwell equations which are Galilei covariant, namely the \emph{magnetic limit} and 
the \emph{electric limit}. Both of them are not of ``radiative'' character which would allow a physically non-trivial quantum field version.}  With the help of $U_{0,1}$ and by tensoring we construct ``Fock space'' $\mathcal{H}_{F,0}$  and the respective representation $U_0 = \oplus_{N=0}^{\infty} \big(U_{0,1}^{\otimes N} \big)_{S}$ of the group $G$ acting in it. 

Recall that in case of non-zero mass $m = \lambda \neq 0$ we would have instead of
(\ref{0m-gal})
\begin{multline}\label{m-gal}
\Psi(x) = (2\pi)^{-3/2} \int 
\limits_{\textrm{sub($\lambda$)}} \hat \Psi(p) e^{i({\bf px} - p_0 t)} d\nu_\lambda(p)\\
= (2\pi)^{-3/2} \int \hat \Psi({\bf p}) e^{i({\bf px} - \epsilon_{\bf p} t)} d^3{\bf p}, \\ \textrm{where} \,\,\, \epsilon_{\bf p} = \frac{{\bf p}^2}{2\lambda}.
\end{multline}
with an irreducible representation of (the double cover) of $SO(3)$ acting in the ``little Hilbert space'' $\mathfrak{h}$ (and of course with the representation of the Bargmann extension $G$ not degenerating to an irreducible representation of the ordinary inhomogeneous Galilean group.)
         
Thus the ``Fock space'' $\mathcal{H}$ of the composed free fields is equal to 
\[
\mathcal{H} = \mathcal{H}_{F,1} \otimes \mathcal{H}_{F,0},
\]
with the representation 
\[
T = U_1 \otimes U_0
\]
of the Bargmann central extension $G$ of the inhomogeneous Galilean group acting in it. 

We consider now the decomposition of the representation $T$ of $G$ into irreducible sub representations with respect to the spectrum 
sp$(P_0, \ldots, P_3, P_4)$, i.e. using decomposition of $\mathcal{H}$ with respect to energy, momentum and mass operators $P_0, \ldots, P_3, P_4$ as in the formulas 
(\ref{dec}), (\ref{decom}). It is now tempting to compare this decomposition of 
$T$ into irreducible representations with the decomposition (\ref{dec}) and (\ref{decom}) of the regular representation $T$ on $\mathcal{M}$ and 
treat it as an abstract ``inverse Fourier transform'' on a spectrally defined
manifold (given by spectral triple) diagonalizing $P_0, \ldots, P_3, P_4$, whose inverse,
i.e. direct transform, will diagonalize the algebra $\mathcal{A}$ dual to that
generated by $P_0, \ldots, P_3, P_4$ (of course with a multiplicity on both sides a priori possible: in the spectrum of sp$(P_0, \ldots, P_3, P_4)$ and in 
sp$\mathcal{A}$, i.e. with generally reducible 
$(\mathcal{A}, D, \mathcal{H})$).\footnote{Note that in 
sp$(P_0, \ldots, P_3, P_4)$ we will have the discretely separated sub manifold sub$(m, \alpha)$ corresponding to the one-particle subspace 
$\mathcal{H}_{m, \alpha} \subset \mathcal{H}$.} 
In other words we expect that the transforms (\ref{m-gal}) and
(\ref{0m-gal}) are respective restrictions of the abstract 
``inverse Fourier transform'' (over the full sub$(P_0, \ldots P_3, P_4)$)
to the sub manifolds ${\bf p}^2 = 2\lambda p_0$ and ${\bf p} = 0, m = \lambda = 0$.
Confirmation of this would give us the undeformed spectral triple $(\mathcal{A}, D, \mathcal{H})$ corresponding to the non relativistic free field in question, composed of the free particles of mass $m$ and spin $\alpha$, accompanied with the pure ``admixture''
component field with no physical quantum particles but plying a crucial role in spectral reconstructing of spacetime from the free quantum field. 

This task is reduced mainly
to decomposition of tensor product representations 
$U_{m, \alpha} \otimes U_{m, \alpha}$, $U_{m, \alpha} \otimes U_{0, 1}$, $U_{0, 1} \otimes U_{0, 1}$ 
into irreducible sub representations. We will not go into further details of this example 
as we are not primarily interesting in the Galilean spacetime.

Thus we suggest that (\ref{m-gal}) -- which is not Fourier transform
in the sense of harmonic analysis -- used in non-relativistic quantum field theory in reconstructing one-particle wave functions\footnote{With well defined locality properties.}, is a restriction of the true Fourier transform involving all irreducible sub representations of the representation $T$ acting in the composed Fock space of free fields.

\vspace{0.5cm}

{\bf Remark 3} It should be stressed that it is non relativistic quantum field
theory which shows the necessity of using the Bargmann central extension
of the Galilean group instead of the Galilean group itself. In the ordinary non relativistic quantum mechanics, in that cases, which
involve just one irreducible representation, 
it is not visible -- in that case we have the equivalence between unitary representations of the Bargmann extension and ray representations of the Galilean group. Indeed although it is the Bargmann extension which acts in the Fock space of a non relativistic field, the assumption that we have a ray (up-to-phase) representation of the inhomogeneous Galilean group in the Fock space would be too strong; although it is sometimes proposed as a tentative axiom for a consistently Galilean quantum field theory, compare e.g. \cite{levy2}.\footnote{It could be already seen in the most elementary case of the
``second quantized'' Schr\"odinger equation.   If such a ray representation existed
in the Fock space, application of the Bargmann theory \cite{bar} would give
us the result that the mass operator has a fixed spectral value all
over the Fock space, corresponding to the value of the parameter (parametrising
inequivalent exponents of ray representations of the Galilean group) which corresponds to the assumed ray representation in the Fock space. 
On the other hand the two-particle subspace is the (symmetrized/antisymmetrized) tensor product of one-particle states. In each of the one-particle subspaces there does act a ray (up to a phase) representation of the inhomogeneous 
Galilean group. Now action of the up-to-phase-representation on the tensor product immediately gives the exponent of the representation in the tensor product equal to the sum of the exponents of the representations in one-particle states. Thus the spectral value of the mass operator in the two-particle subspace is double the value on one-particle subspace. Thus the spectral values of the mass operator cannot be constant all over the whole Fock space. Only the assumption that we have a unitary representation of the Bargmann extension -- explicitly asserted in the cited paper of L\'evy-Leblond \cite{levy2} -- seems to be the correct substitute for Galilean covariance, supported experimentally by non relativistic quantum field theory models.

The quantum field $\Phi (t,{\bf x})$ transforms under the elements 
$(r,0) \in G = G' \times \mathbb{R}$ of Bargmann extension $G$ representing inhomogeneous Galilean group $G'$ with the phase factor at the front exactly the same as that in the transformation law for one-particle wave function , with the value for the mass parameter just the same as for one particle states. Here the mass operator and its spectral values behave like a generalized charge, but field operators
mix the super selection sectors given by this charge.}  

\vspace{0.5cm}

{\bf Example 2} Let $\mathcal{M}$ be the Minkowski spacetime. Consider the free
quantum Klein-Gordon field of mass $m$. By tensoring we may construct 
the field from the irreducible representation $U_{m,0}$ of the double cover $G$ 
of the Poincar\'e group, corresponding to the mass $m$ and spin zero, constructed after
Wigner, as has already been indicated in the Example 1 above with not entirely
controlled assumptions about the representation mentioned to above. This time 
Bargmann extension degenerates just to the double cover, so there is no
extra operator\footnote{I.e. generator of the center $Z$ of the extension, as the central extension is trivial here, with $Z = \{1\}$, and contracts locally to the initial group -- the Poincar\'e group.} $P_4$ which would have to be added to the generators of translations $P_0, \ldots, P_3$ -- the energy, momentum operators in order to compose a maximal commuting algebra of generators (or their functions) of the 
group $G$. Thus we decompose a given representation of $G$ in $\mathcal{H}$ with respect
to the joint spectrum of $P_0, \ldots, P_3$, i.e. we decompose $\mathcal{H}$ 
\[
\mathcal{H} = \int \limits_{\textrm{sp$(P_0, \ldots, P_3)$}} \mathcal{H}_pd\nu(p)
\]
into a direct integral of generalized common proper subspaces of
$P_0, \ldots, P_3$ (after Wigner, just like in the preceding example).
After restricting the continuous integral decomposition to the hyperboloid $p_0^2 - 
{\bf p}^2 = \lambda^2$ of mass $m = \lambda$, i.e. to a sub manifold sub($\lambda$) of 
sp$(P_0, \ldots, P_3)$, and using the Wigner's technique of ``little Hilbert space'' we obtain the irreducible subspace $\mathcal{H}_{\lambda,\alpha}$  corresponding to an irreducible sub representation in the following form 
\[
\mathcal{H}_{\lambda,\alpha} 
= \mathfrak{h} \otimes \mathfrak{H}, \,\,\, \textrm{where} \,\,\, 
\mathfrak{H} = L^2(\textrm{sub($\lambda$}), d\nu_\lambda(p)),
\]
with respect to the measure $d\nu_\lambda(p)$ on sub($\lambda$) induced by the spectral measure $d\nu(p)$; and $\mathfrak{h}$ is the ``little Hilbert space'' with the ``little
group''\footnote{I.e. double cover of $SO(3)$.} SU(2) acting irreducibly in 
$\mathfrak{h}$ if $m \neq 0$. Thus $\mathfrak{h}$ depends on the 
irreducible representation of (the double cover of) the rotation group 
(for $m \neq 0$), and thus on the
chosen spin $\alpha$. Let us denote the irreducible subspace $\mathcal{H}_{\lambda,\alpha}$ by $\mathcal{H}_{m,\alpha}$ and the irreducible representation of $G$ 
acting in it by $U_{m,\alpha}$. In case of zero mass representation the 
``little group'' is equal to symmetry group $E_2$ of the (two-dimensional) Euclidean plane. 

Thus the elements of the irreducible subspace $\mathcal{H}_\lambda$ are $\mathfrak{h}$-valued functions $p \mapsto \hat \Psi(p)$ on sub($\lambda$) of the momentum $p$.

In case of $m \neq 0$ we construct in analogy to the non relativistic case
a wave function, i.e. kind of a restricted Fourier transform
\begin{multline}\label{m-poi}
\Psi(x) = (2\pi)^{-3/2} \int 
\limits_{\textrm{sub($\lambda$)}} \hat \Psi(p) e^{i({\bf px} - p_0 t)} d\nu_\lambda(p)\\
= (2\pi)^{-3/2} \int \hat \Psi(\epsilon_{\bf p},{\bf p}) e^{i({\bf px} 
- \epsilon_{\bf p} t)} \frac{d^3{\bf p}}{2\epsilon_{\bf p}}, \\ \textrm{where} \,\,\, \epsilon_{\bf p} = ({\bf p}^2 + m^2)^{1/2}.
\end{multline}    
In this case of the doubly covered Poincar\'e group $G$ and the Minkowski
spacetime $\mathcal{M}$ to the restricted transform (\ref{m-poi})
there does not exist reasonable restriction in the spacetime variable,
i.e. the corresponding sub manifold sub$(\mu)$\footnote{I.e. analogue of simultaneity hyperplane in Galilean spacetime.} of $\mathcal{M}$ such that (\ref{m-poi})
would give us a unitary mapping between $L^2(\textrm{sub$(\lambda)$}, d\nu_\lambda(p))$
and $L^2(\textrm{sub$(\mu)$}, d\upsilon_\mu(p))$. In particular the positive energy 
wave functions do not compose any complete system in restricting say to the hyperplane
$t = const$ of a Lorentz frame. However, as Newton-Wigner analysis reveals, 
in case of non zero mass the wave function gives a rough localization within
$t = const$ hyperplane with the ambiguity caused by its Lorentz-non-invariance not exceeding the Compton wave length. We thus not have any well defined position operator
in the ordinary quantum mechanical sense as in the non-relativistic case, but it is precisely what we expect, as we view (\ref{m-poi}) as a restriction of a true Fourier
transform on $\mathcal{M}$, and on the other hand restrictions of Fourier transform 
do not provide in  general any well defined unitary transforms. 

Note that in case of higher spin it is necessary to use a (Foldy-Wouthuysen) 
transformed $M(p)\hat \Psi(p)$ Wigner 
$\mathfrak{h}$-valued momentum function $\hat \Psi(p)$ 
with the $p$-dependent matrix $M(p)$ in order to achieve a transparent locality 
analysis with local transformation law for wave functions; which suggest that 
in this case either using a non-trivial functions of the operators $P_0, \ldots, P_3$ is more desirable than the energy-momentum operators themselves or slightly modified Wigner representations diagonalizing $P_0, \ldots, P_3$. We will chose the second possibility in the Example 3 below.     

Using $U_{m,0}$ we construct by tensoring the free quantum Klein-Gordon field in the Fock space $\mathcal{H}_{F,1}$ together with the respective unitary representation 
$U_1$ in $\mathcal{H}_{F,1}$. 
  
Altough for massless particles the concept of localization is not
appropriate we add to this non-zero mass particle field the free electromagnetic
field in order to elucidate connection to the full harmonic analysis on 
$\mathcal{M}$, as in the preceding example. And then we form a composed 
non-interacting free fields. The non-trivial ingredient
is the construction of the free electromagnetic field from an appropriately chosen
irreducible representation of $G$ acting in a Krein space $\mathcal{H}$; and thus in making the correct guess in constructing both $\mathcal{H}$ and the irreducible ``unitary'' representation $U_{0,1}$ in $\mathcal{H}$, suitable for the construction
of the free field by tensoring. This task however has been already solved by 
{\L}opusza\'nski \cite{lop1,lop2}. Because his works are in Polish and German, let us briefly sketch the idea of {\L}opusza\'nski's  \cite{lop1, lop2} construction. In fact we have already followed him in the preceding Example. He proceeds after Wigner as far as possible, i.e. assuming that we have abstractly given representation of $G$ in a space $\mathcal{H}$ as if it was a Hilbert space and the representation unitary. Moreover he assumes the representation to be sufficiently regular, as if it was a ``regular representation'' acting on the space-time manifold, in particular with a Lebesgue-type spectral measure in the joint spectrum of translation generators. As we have already stressed above these not entirely controlled assumptions are not specific for the {\L}opusza\'nski construction but pertain to the very heart of the construction of one-particle wave functions in QFT.  He proceeds by a perhaps boringly familiar 
by now procedure and decompose $\mathcal{H}$ with respect to $P_0, \ldots, P_3$ showing then, that the  inverse Fourier transform ``restricted'' to the sub manifold 
sub($\lambda =0$) (i.e. in case of zero mass $m = \lambda = 0$): 
\begin{multline}\label{0m-poi}
\Psi(x) = (2\pi)^{-3/2} \int 
\limits_{\textrm{sub($\lambda$)}} \hat \Psi(p) e^{i({\bf px} - p_0 t)} d\nu_\lambda(p)\\
= (2\pi)^{-3/2} \int \hat \Psi(\epsilon_{\bf p},{\bf p}) e^{i({\bf px} 
- \epsilon_{\bf p} t)} \frac{d^3{\bf p}}{2\epsilon_{\bf p}}, \\ \textrm{where} \,\,\, \epsilon_{\bf p} = |{\bf p}|.
\end{multline}    
is a four-vector on the spacetime $\mathcal{M}$ of helicity 1 with the irreducible representation space $\mathcal{H}_{\lambda = 0, 1}$ allowing to a non-trivial subspace of positive-norm vectors if and only if the (irreducible) space 
\[
\mathcal{H}_{\lambda = 0, 1} 
= \mathfrak{h} \otimes \mathfrak{H}, \,\,\, \textrm{where} \,\,\, 
\mathfrak{H} = L^2(\textrm{sub($\lambda$}), d\nu_{\lambda = 0}(p)),
\]
is a Krein space, or more precisely if the ``little space'' $\mathfrak{h}$ 
is a four dimensional Krein space. It should be stressed that by
the nature of the construction $\mathcal{H}_{\lambda = 0, 1}$ 
is indeed a Krein space as $\mathfrak{H} = L^2(\textrm{sub($\lambda$}$ 
is an ordinary Hilbert space. Similarly the proper subspaces of ``unitary representors'' of translations in the {\L}opusza\'nski representation $U_{0,1}$ are by construction non-degenerate in the sense of \cite{Bog}, with unimodular generalized proper values, similarly their generators $P_0, \ldots, P_3$ with $\mathcal{H}_{\lambda = 0, 1}$ already decomposed into generalized proper vectors of $P_0, \ldots, P_3$. This greatly simplifies decomposition of $U_{0,1} \otimes U_{0,1}$ into irreducible representations and the tensor product space $\mathcal{H}_{\lambda = 0, 1} \otimes \mathcal{H}_{\lambda = 0, 1}$ into irreducible subspaces. It should be stressed that in general Krein space this is far not the case for an abstractly given ``unitary'' representation. In particular
decomposition theory is by no means automatic for such a general ``representation'' in a Krein space (similar comments could be said concerning the first Example). 
We construct Fock space $\mathcal{H}_{F,0}$ of the free electromagnetic field 
by tensoring the one-particle irreducible space $\mathcal{K}_{\lambda = 0, 1}$  
of {\L}opusza\'nski representation $U_{0,1}$, together with the ``unitary'' representation $U_0 = \oplus_{N=0}^{\infty} \big(U_{0,1}^{\otimes N} \big)_{S}$ of the group $G$ acting in it. And exactly as before we have the representation 
\[
T = U_1 \otimes U_0
\]
 of $G$ in the composed Fock space 
\[
\mathcal{H} = \mathcal{H}_{F,1} \otimes \mathcal{H}_{F,0},
\]
of the composed Klein-Gordon and electromagnetic free fields.

Now as before we decompose the representation $T$ using the decomposition
of the Fock space with respect to $P_0, \ldots, P_3$ and compare the decomposition
with (\ref{dec}) and (\ref{decom}). Note that in 
sp$(P_0, \ldots, P_3)$ we will have the discretely separated 
sub manifold sub$(m, \alpha = 0)$ corresponding to the one-particle subspace 
$\mathcal{H}_{m, \alpha =0} \subset \mathcal{H}$. It is now tempting to compare
(\ref{m-poi}) to the inverse Fourier transform into irreducible 
sub representations of $T$
restricted to the sub manifold sub$(\lambda = m, \alpha = 0) =
\textrm{sub}(m, \alpha = 0)$ on a bona fide spacetime manifold. 

\vspace{0.5cm}

{\bf Example 3} Consider now a QFT with a mass gap
and non-empty discrete part of the spectrum of the mass operator 
$M = (P_\mu P^\mu)^{1/2}$, together with the Haag-Ruelle collision theory, 
compare \cite{Haag}, Chap. II.3-II.4 and references therein.
In the Hilbert space $\mathcal{H}_1$ of the theory we have a unitary representation
$U_1$ of the double covering $G$ of the Poincar\'e group. 
We add to the fields of theory the free uncoupled electromagnetic field, which from the
physical point of view would be redundant, but again is necessary in connection to
harmonic analysis on $\mathcal{M}$. We construct the free electromagnetic field
with the help of the {\L}opusz\'nski representation in the Krein ``Fock'' space
$\mathcal{H}_{F,0}$, and then construct the representation $T = U_1 \otimes U_0$ of    
of $G$ in the Krein space $\mathcal{H} = \mathcal{H}_{F,0} \otimes \mathcal{H}_1$ of the composed system of fields. If the initial QFT is asymptotically complete, then the 
Hilbert space $\mathcal{H}_1$ can be realized as the tensor product
$\Pi_i^\otimes \mathcal{H}_{F,i}$ of Fock spaces composed of asymptotic particle states
allowed by the theory, with the representation 
\[
U_1 = \Pi_i^\otimes U_{F,i},
\]
where 
\[
U_{F,i} = \Sigma_N^\oplus \big(U_{m.,\alpha_i}^{\otimes N} \big)_{S,A},
\]
similarly as in a system of free uncoupled fields each corresponding
to the particle species $i$.

Now in the reduction of $T$ some irreducible representations 
should occur with a discrete weight, namely those corresponding to the discrete
part of the spectrum of the mass operator $(P_\mu P^\mu)^{1/2}$, 
and thus some sub manifolds sub$(\lambda)$:  
$p_0^2 - {\bf p}^2 = \lambda^2$ of mass $m = \lambda$ of the joint
spectrum sp$(P_0, \ldots, P_3)$, corresponding to the discrete proper values $m$
of the mass operator have discrete weight. Using the irreducible 
representation corresponding to such a discrete sub manifold sub$(\lambda)$
we construct by application of the ''restricted'' inverse Fourier transform 
(\ref{m-poi}) a single particle wave function of mass $m$ and spin determined by the representation.        

\vspace{0.5cm}

Summing up: we construct explicitly the representation $T$ of the double cover $G$ of the Poincar\'e group acting in the Krein space $\mathcal{H}$ of free QED fields. 
Because we construct the free electromagnetic field using the {\L}opusza\'nski representation as in the previous two Examples, then the representation $U_0$ of $G$ in the Fock space of electromagnetic field is already at hand. It remains to construct explicitly the representation $U_1$ of $G$ acting in the Fock space of the free Dirac field. To this end it is sufficient to give the explicit formula for the sub representation acting in one-particle electron space and in one-particle positron space. But those sub representations have explicit form both being reducible direct sums of the conjugate Majorana representations (of course we mean in the momentum space representation in which $P_-, \ldots, P_3$ are diagonal).  
Having the representation $T = U_1 \otimes U_0$ in $\mathcal{H}$ we treat 
it as a regular representation (up to a possible multiplicity) acting on a bona fide 
spacetime manifold. Namely we decompose $T$ into irreducible representations, diagonalizing\footnote{I.e. in which the operators $P_0, \ldots, P_3$ are ``diagonal'', and have the form of multiplication operators.} $P_0, \ldots P_3$ and treat this decomposition as an ''inverse Fourier transform'' (up to multiplicity) on the bona fide manifold, whose inverse, i.e. direct transform, should diagonalize the algebra $\mathcal{A}$ of its coordinates.  

This task however reduces to the problem of decomposing tensor product
representations of two representations coming from the following set of representations: the two  Majorana irreducible representations mentioned to above and the {\L}opusza\'nski representation. Problems of this type are rather extensively examined, compare for example the series of works, \cite{nai1}-\cite{nai3} and \cite{tat}-\cite{puk}, where  decomposition of tensor products of irreducible unitary representations of the Lorentz group and of the Poincar\'e group has been examined. The circumstance however that we are interested in concrete Wigner-type representations diagonalizing $P_0, \ldots, P_3$ and not just in decompositions which are unitarily equivalent to the concrete tensor products of the given irreducible representations but rather in decompositions themselves into irreducible representations which diagonalize $P_0, \ldots P_3$, we prefer to perform computations separately in our case in order to solve our
problem.    

\vspace{0.5cm}

{\bf Remark 4} From the point of view presented here the algebra of operators 
$\mathcal{A}$ gives the spacetime points as its spectrum (its irreducible sub representations) with possible multiplicity, as classical non superposing parameters.
Elements of $\mathcal{A}$ cannot be treated as quantum mechanical operators. It is only an accident coming from the specific group structure and the spacetime 
manifold $\mathcal{M}$ that one can restrict the inverse Fourier transform on 
sp$(P_0, \ldots P_3)$ to a sub manifold corresponding to an irreducible representation of one-particle states obtaining a unitary operator on
functions restricted to simultaneity hyperplanes with a well defined position operators
for each time separately. Thus the existence of the ordinary 
non relativistic quantum mechanics is an immediate consequence of this accidental
structure of the Galilean group and the Galilean spacetime which allows such a restriction of the full spacetime Fourier transform. Quantum field theory is more fundamental even in non relativistic case, and even in non relativistic quantum field theory the spacetime coordinates are classical (non-superposing) parameters. Therefore the ordinary non-relativistic quantum theory is emergent in relation to 
(non-relativistic) quantum field theory and is completely needles for the purposes of quantum field theory.  In particular there is no need for the position operator if one uses spectral construction of spacetime manifold
provided one uses in addition the geometric Haag interpretation. In particular that there is no well defined position operator in relativistic theory is what we should expect from this point of view. Indeed using the spectral construction of spacetime and the geometric interpretation of Haag the Newton-Wigner analysis accompanying massive particles and the theory with mass gap, which serves us a substitute for the (now absent) position operator is completely unnecessary. Therefore the spectral construction of spacetime allows us to generalize the theory so as to embrace theories with zero mass particles (no mass gap) such as QED.

\subsection{Multiplicity of the algebra generated by $P_0, \ldots, P_3$ or 
respectively by $P_0, \ldots, P_3, P_4$ in the Fock space $\mathcal{H}_{F,1}$}

Here we examine only qualitatively non-cyclicity (spectral multiplicity) 
of the algebra generated by the operators 
$P_0, \ldots, P_3$ (eventually $P_0, \ldots, P_3, P_4$) in restriction to 
the Fock space\footnote{It may be considered as an actual subspace of 
$\mathcal{H} = \mathcal{H}_{F,1} \otimes \mathcal{H}_{F,0}$ corresponding
to (photon vacuum)$\otimes \mathcal{H}_{F,1}$.} $\mathcal{H}_{F,1}$ 
composed of non-zero mass particles of Examples 1-2 and eventually
for the free theory underlying QED. 

Let us suppose for simplicity that the spin $\alpha = 0$. For the two-particle 
subspace $\big(\mathcal{H}_{m,\alpha =0} \otimes \mathcal{H}_{m,\alpha =0}\big)_S$ the states are symmetrized tensor products of Wigner's states, i.e. (symmetrized) complex valued functions $(p_1,p_2) \mapsto \Psi(p_1,p_2)$ on $\operatorname{sub}(\lambda = m) \times \operatorname{sub}(\lambda = m)$, where $\operatorname{sub}(\lambda)$ is the submanifold
in $\operatorname{sp}(P_0, \ldots, P_3)$ (resp. in 
$\operatorname{sp}(P_0, \ldots, P_3, P_4)$) corresponding to the irreducible subspace 
$\mathcal{H}_{m,\alpha =0}$ of the subrepresentation $U_{m,\alpha=0}$. The scalar product
is 
\[
(\Phi, \Psi) = \int d\nu_{\lambda = m}(p_1) \, d\nu_{\lambda = m}(p_2) \,
\overline{\Phi(p_1, p_2)} 
\Psi(p_1,p_2)
\]
together with the transformation law $(U_{m,0} \otimes U_{m,0} \big)_S$. From what we know of two particle systems combined of one-particle constituents it follows that
the masses $m'$ of the two-particle states vary from $m+m$ to infinity (resp. $m'=m+m$ in Galilean case), as the relative momentum of the particles varies in magnitude from zero to infinity. Better: it follows from the fact that 
$P_i|_{\mathcal{H}_{m,0} \otimes \mathcal{H}_{m,0}} \big( \Phi_1 \otimes \Phi_2 \big) 
= P_i \Phi_1 \otimes \Phi_2 + \Phi_1 \otimes P_i\Phi_2$ and from the relation between 
${\bf p}$ and $p_0$ on $\operatorname{sub}(\lambda = m)$ corresponding to the irreducible subspace $\mathcal{H}_{m,0}$. The amplitude for ${\bf p}_1 + {\bf p}_2 = 0$
still depends on the relative momentum, so that under rotations all orbital angular momenta $l$ will in general participate. That means that in decomposition of the
two-particle subspace $\mathcal{H}_{m,0} \otimes \mathcal{H}_{m,0}$ into irreducible
subspaces there will participate all irrreducible subspaces $\mathcal{H}_{m',\alpha}$
with $m' \geq m + m$  (resp. just $m'=m+m$ in Galilean case) 
and with all integer spins $\alpha$ correponding to
the irreducible representations $U_{m', \alpha}$. And from what we know about 
two-particle systems the two quantum numbers: $m'$ and $\alpha = l$ uniquelly
define the action of generators of the symmetry group in quection (double covering of Poincar\'e group or respectively the central Bargmann extension of inhomogeneous Galilean group); in other words $m', \alpha$ and amplitude for three-momenta uniquelly
define a two-particle state. Thus in decomposition of 
$(U_{m,0} \otimes U_{m,0} \big)_S$ into irreducible Wigner-type subrepresentations
$U_{m',\alpha}$ every $U_{m',\alpha}$ enters with multiplicity 
one.\footnote{Compare \cite{wig}, where the argumet is used on page 25.} 
Thus the multiplicity of the spectrum of $P_0, \ldots, P_3$ 
(resp. $P_0, \ldots, P_3, P_4$) after restriction to the subspace $\mathcal{H}_{F,1}$ is purely discrete. Of course this is only qualitative argument, which should be 
read of from the decomposition of $U_{m,\alpha} \otimes U_{m,\alpha}$ into irreducible representations $U_{m,\alpha}$ diagonalizing the generators $P_i$ --
a task we postpone to another occasion.

Because of this purely discerete character of multiplicity of $P_0, \ldots, P_3$ (resp. $P_0, \ldots, P_3, P_4$) we see in case of the Galilean case of Example 1 that for the ordinary riemannian spectral triple $(\mathcal{A}_{qu}, D_{qu}, \mathcal{H}/\mathcal{H}_0)$, obtained from $(\mathcal{A}, D, \mathcal{H})$ by quotiening out the closed subspace $\mathcal{H}_0$ of zero norm vectors, the quotient algebra 
$\mathcal{A}_{qu}$ shlould have discrete multiplicity in the ordinary Hilbert space 
$\mathcal{H}/\mathcal{H}_0$; irrespectevily of the character of multiplicity of the algebra $\mathcal{A}$  in $\mathcal{H}$, coming from the ``admixture'' photon component. Similarly in the relativistic case of the free theory underlying QED we expect
the possibly highly non-trivial multiplicity of $\mathcal{A}$ to cease in 
passing to the ordinary riemannian spectral triple 
$(\mathcal{A}_{\mathfrak{J}}, D_{\mathfrak{J}}, \mathcal{H}_{\mathfrak{J}})$. 

Of course in case of highly nontrivial multiplicity of $\mathcal{A}_{\mathfrak{J}}$
application of Fedosov method would be difficult because it woud be difficult
to treat the operators in $\mathcal{H}_{\mathfrak{J}}$ as operators on bona fide (spectrally defined) manifold 
$(\mathcal{A}_{\mathfrak{J}}, D_{\mathfrak{J}}, \mathcal{H}_{\mathfrak{J}})$, 
with well defined abstract symbol calculus. Note also that after deformation,
when the  interaction is switched on the purely discrete multiplicity character should be essentially preserved, compare the multiplicity assumption 3. of \cite{wig} on page 29, and the arguments supporting it given there.

\section{Time's Arrow for Non-superposing Quantities}\label{time}

Vector fields (e. g. the vector field corresponding to time evolution) on an ordinary manifold correspond canonically to one-parameter groups of automorphisms of the algebra of smooth functions on the manifold (e. g. the one parameter group of time automorphisms). The non-commutative multiplication in the algebra of space-time coordinates has the mathematical consequence that the "non-commutative transformations" corresponding to a vector field are not automorphisms of the algebra (a phenomenon connected to Morita equivalence) and do not form any group in the ordinary sense in general. There are several competitive structures which have to replace the ordinary group (the so called \emph{quantum group} is one of the main candidates\footnote{Until recently it was widely believed that quantum groups do not fit into the spectral triple format. Quite recent works show that the two formalisms may be reconciled. Let us cite the breakthrough papers only: \cite{Chak, Conn, Sui1, Sui2}}) but it is beyond doubt that in general the group property ensuring the existence of the inverse transformation among the "non-commutative transformations" for any "non-commutative transformation" (e. g. ensuring the existence of the backward time evolution "$-t$"  for every time evolution "$+t$") is not fulfilled in general. This is the case for example for quantum groups. However the possibility that some classical parameters corresponding to spectra of some commutative sub-algebras of spacetime coordinates  are acted on by the quantum group (determining say the time evolution) as by an ordinary one-parameter group in not \emph{a priori} excluded; in other words: besides the classical parameters evolving non-deterministically, there could in principle exist parameters evolving deterministically. To explain this let us consider a model. Because the full theory involves extremely complicated computational machinery, and moreover one of its most fundamental ingredients is not explicitly constructed, i. e. the operator $D$, we are forced to consider a very simplified (even oversimplified) model. Namely we consider quantum fields in two-dimensional spacetime, which are completely integrable, constructed by Faddeev and his school, such e. g. as the quantized nonlinear Schr\"odinger or sine-Gordon equation. They are constructed from the classical inverse scattering transform, just by replacement of the "classical" fields in the monodromy matrix with point-like operator valued distributions, thus obtaining the quantum monodromy matrix $\intercal(\lambda)$, compare the monograph of Korepin, Bogoliubov and Izergin \cite{KorBogIze}, and utilizing the normal ordering (Wick theorem). Let us remind that in such models (two dimensional spacetime) renormalization is finite \cite{GliJaf} (no Haag's theorem) so that the interacting fields may be represented in the Fock space along with free fields, and there is no necessity in smearing them out over open sets of full dimension. The distributions in the monodromy matrix $\intercal(\lambda)$ so obtained, which in general are only sesquilinear forms on a dense subset $H_0$ of the Hilbert space (here the Fock space),  can  moreover be multiplied (Wick theorem applicable) on the dense subset in this simplified situation. Here the dense subset $H_0$ is obtained when acting on the Fock vacuum state by all the polynomials in elements of the second column of the monodromy matrix. Thus we obtain a linear representation (quite singular from the analytic point of view) of the set of linear operators, i. e. the monodromy matrix elements, on the linear subspace $H_0$. As shown in \cite{KorBogIze} the construction of the monodromy matrix $\intercal(\lambda)$ is equivalent with determination of the time evolution (e. g. in the case of "second-quantized" nonlinear Schr\"odinger equation, it is equivalent to the Bethe Ansatz). Let us stop for a moment at the pure linear-algebraic level of the mentioned representation in the linear space $H_0$ without any care for analytic subtleties in assuring a strict mathematically well defined relationship to the Fock space, keeping in mind only the formal analogy to the Fock space inscribed in the construction of the representation. This is what mathematicians actually did when inventing quantum group. Namely the algebra generated by the elements of the monodromy matrix is from the pure algebraic point of view an algebraic quantum group in the sense of Manin\footnote{In fact \emph{quantum groups} were invited by Drinfeld \cite{Dri}, who placed the algebra into the category of specific bi-algebras with adequately defined structure embracing the algebras of  smooth functions on Lie groups with the fully fledged adequately rigid topological structures, generalizing the properties of algebras of smooth functions on Lie groups,  asserting non-triviality of the theory of representations of the object. Prof. S. L. Woronowicz introduced the topological structure along the C*-algebra format and extended the Peter-Weyl theory on the quantum compact groups. Further analytic structures, as e. g. differential structure along the spectral triple format was invited in the papers cited in 31 footnote. However the topological and analytical structures invited thereafter have no clear connection to the whole analytic structure of the initial physical situation (Faddeev models).}     \cite{Manin1}. Thus in the analysis of the algebra (quantum group) we follow mathematicians for a while in order to make clear our motivation for the last task of the proposal. From the commutation relations of the algebra\footnote{\emph{I. e.} commutation relations of the monodromy matrix elements.} (quantum group) it follows that  it coacts on the algebra generated by the  first column of the monodromy matrix\footnote{Compare \cite{KorBogIze}, p. 47.}. The later corresponds formally to the algebra of annihilators with adjoined unit, and thus correspond to our spacetime algebra, via the correspondence between fields and local algebras, which is assured in this completely integrable case. In general the quantum group so constructed is a Yangian, whose structure is still quite complicated. This is the case for the nonlinear Schr\"odinger and sine-Gordon models at least. In particular the Yang-Baxter matrix with parameter $R(q_{1},q_{2})$ corresponding to it has a pole at $q_1 = q_2$. Therefore we go further in our mathematical simplifications and assume that we have such a model (if there exits such and is still reasonable) whose Yang-Baxter matrix is not singular at $q_1 = q_2$, so that we can assume that $R$ is a function of one parameter $q$ only. Then we may use the root of unity phenomenon, investigated and generally described mostly by Lusztig \cite{Lusz1, Lusz2}. Namely, if $q$ is a primitive root of unity of odd degree, then quantum groups corresponding to Yang-Baxter matrices with one parameter $q$ contain a "big" commutative sub-algebras and the structures of the quantum groups generate natural ordinary group structures on these sub-algebras and the actions of the quantum groups on their uniform spaces induce ordinary group actions on the spectra of the commutative sub-algebras. Perhaps the Manin group $GL_{q}(2)$ co-acting on the algebra of the Manin plane is the simplest illustration of the phenomenon, compare e. g. \cite{Manin2}, pp. 151-153. In this case the mentioned sub-algebras lie in the center (of the corresponding algebras). Because on the other hand the algebras "of  functions" of these quantum groups and of their uniform spaces are not in general Morita equivalent to commutative algebras, even for $q$ equal to a primitive root of unity, then their actions on the spectra of commutative sub-algebras is not in general equivalent to ordinary group actions. In particular neither the algebra "of functions" of $GL_{q}(2)$ nor the algebra of the Manin plane are Morita equivalent to commutative algebras, even if $q$ is equal to a primitive root of unity \cite{RichSol}.            
    
\vspace*{0.5cm}

\begin{tabular}{|p{11.2cm}|}\hline
Thus we arrive at the fourth task of our proposal: to investigate more deeply the analytic properties of the linear representation of the quantum monodromy matrix 
$\intercal(\lambda)$  on the dense subset $H_0$ of the Fock space, given in \cite{KorBogIze}. Then incorporating the relationship between point-like fields and local algebras (as developed in the following papers: \cite{BorZim, DriFro, BorYng}) try to carry the quantum group structure and their action on the corresponding spacetime algebra of bounded operators. The goal is to convert the formal argument demonstrated above into an actual. \\ \hline
\end{tabular}

\section{Our Hypothesis and the Onsager Principle}\label{onsager}

In our proposal the tentative hypothesis of Sect. \ref{hyp} plays a crucial
role. It says that the essential point of DHR analysis of generalized 
charges may be extend  so as to embrace all classical 
(in the sense: non-superposing) quantities. That is, we assume that all non-superposing (``classical'' so to say) quantities should be decomposition parameters of a distinguished sub algebra of the algebra of field operators corresponding to the classical quantities(s) in question. We have applied it to the algebra of spacetime coordinates $\mathcal{A}$ in
our proposal. Thus it is interesting for us if indeed DHR analysis
may be so extended, and try to find some physically verifiable consequences of such 
extension. Here we examine qualitatively such extension
outside the realm of high energy physics.
 
Namely suppose we have a complicated system, i.e with quite a huge number of degrees of freedom treated as non-superposing parameters, just such as we encounter in classical statistical mechanics. Now let us assume (assumption which of course may a priori be false) that DHR analysis is applicable to these non-superposing (huge in number)
parameters. Let us try go as far as possible with this assumption in deriving some qualitative at least physical consequences. Now depending on the specific character of the non-superposing parameters, the corresponding algebra may be (Morita equivalent to) a commutative algebra or not. As we are forced to remain at this general qualitative 
level we may only infer a very general conclusion, namely, that if the algebra is essentially non-commutative\footnote{Which by no means stands in contradiction to the 
classical superposition-less character of the parameters numbering the selection sectors
in the representation space of the corresponding algebra, as has already been mentioned above in Sect. \ref{hyp}.}, then the one-parameter group of time transformations would have to be modified into a quantum group action with the time reversal law broken.
On the other hand we have a very deep (perhaps completely forgotten by now) recognizion of Sir Isaac Newton, that the multiplication structure of physical quantities with physical dimension should be introduced by tensor quantities which actually
do exist in reality, i.e. quantities which multilinearly depend on them, 
compare \cite{waw} where we explain in details the 
ingenious recognition of Newton.\footnote{It is in XX$^{th}$ century
mathematics where the ingenious idea of Newton was rediscovered in constructing algebras as quotients of the tensor product algebra over a fixed vector space (compare e.g. algebraic theory of quantum groups); it seems that the deep recognizion of Newton has escaped adequate attention of physicists.} From this we can conclude that if the algebra is non commutative and thus time arrow is unavoidable, we should observe essentially non symmetric tensor quantities corresponding to the essentially non-commutative algebra describing the classical parameters. Of course there are some ambiguities on both sides: two Morita equivalent algebras have isomorphic representation spaces, and we have some ambiguities in defining tensor quantities, for example that recognized by Casimir in transport processes, as well as some other much less easy to control in practice in extracting the relevant physical content. Nonetheless we can infer a general rule, that the time arrow should be accompanied by existence of (essentially) non-symmetric material tensors. This conclusion is quite reasonable for at least two reasons. 
 
Let us give the first reason. We have namely the Onsager principle in the transport processes. Namely Onsager \cite{onsager1,onsager2} was able to prove, using tricky methods of Einstein and the Gibbs method, that tensors describing transport phenomena (such as the heat conductivity tensor) should be symmetric whenever we assume ``microscopic irreversibility'' to hold.       
  
The second reason comes from the results of Kac \cite{kac}. Namely he devoted almost all his live in examining equivalence between the stochastic method of Smoluchowski and the 
method of Gibbs. Conclusion he arrived at is presumably negative \cite{kac}: some additional random mechanism in the time evolution law in the Gibbs method is needed in order to recover all the results obtained with the help of master equation. 

This is not the end of the history. We can go somewhat further with our hypothesis
at hand. Namely because the Planck constant is very small in comparison to action
involved in macroscopic processes, we expect that the algebra in question
is ``practically'' commutative with material tensors almost (essentially) symmetric, obtaining the conclusion that the Onsager principle can be fulfilled only approximately. Because the Planck constant is non zero we should observe small deviations from 
that principle. Over one hundred years ago Soret \cite{soret1,soret2} and 
Voight \cite{voigt} had experimentally verified existence of non-symmetric deviations from the Onsager principle obtaining negative results. Possibly a repetition of such experiments with the modest highly sensitive calorimetric tools would not be devoid of reasons.

\vspace*{1cm}

{\bf ACKNOWLEDGEMENTS}

\vspace*{0.5cm}

The author is indebted for helpful discussions to prof. A. Staruszkiewicz and particularly for the discussion in April of 2008 during his visit to INP PAS.  The author would especially like to thank prof. M. Je\.zabek for the warm encouragement
and for the excellent conditions for work at INP PAS where the most part of this proposal has come into being.

\end{document}